\documentclass[a4paper,11pt]{article}
\pdfoutput=1 
\usepackage{jcappub} 
\newcommand{\lsim}{\mbox{\raisebox{-.6ex}{~$\stackrel{<}{\sim}$~}}}
\newcommand{\gsim}{\mbox{\raisebox{-.6ex}{~$\stackrel{>}{\sim}$~}}}
\title{Constraining Non-thermal and Thermal properties of Dark Matter}
\author[a]{P. S. Bhupal Dev,}
\author[b]{Anupam Mazumdar,}
\author[b,c]{and Saleh Qutub}
\affiliation[a]{~Consortium for Fundamental Physics, School of Physics and Astronomy,
University of Manchester, Manchester, M13 9PL, United Kingdom}
\affiliation[b]{~Consortium for Fundamental Physics, Physics Department, Lancaster University, LA1 4YB, United Kingdom}
\affiliation[c]{~Department of Astronomy, King Abdulaziz University, Jeddah 21589, Saudi Arabia}
\abstract{We describe the evolution of Dark Matter (DM) abundance from the very 
onset of its creation from inflaton decay under the assumption of an instantaneous reheating. Based on the initial conditions such as the 
inflaton mass and its decay branching ratio to the DM species, 
the reheating temperature, and the mass and interaction rate of the DM with the thermal bath, the DM particles can either thermalize (fully/partially) with the primordial bath or remain non-thermal throughout their evolution history. In the thermal case, the final abundance is set by the standard freeze-out  mechanism for large annihilation rates, irrespective of the initial conditions. 
For smaller annihilation rates, it can be set by the freeze-in mechanism which also does not depend on the initial abundance, provided it is small to begin with. For even smaller interaction rates, the DM decouples while being non-thermal, and the relic abundance will be essentially set by the initial conditions. 
We put model-independent constraints on the DM mass and annihilation 
rate from over-abundance by exactly solving the relevant Boltzmann equations, and identify the thermal freeze-out, freeze-in and non-thermal regions of the allowed parameter space. We highlight a generic fact that inflaton decay to DM inevitably leads to an 
overclosure of the Universe for a large range of DM parameter space, and thus poses a stringent constraint that must be taken into account while constructing models of DM. For the thermal DM region, we also show the complementary constraints from indirect DM search experiments, Big Bang Nucleosynthesis, Cosmic Microwave Background, Planck measurements, 
and theoretical limits due to the unitarity of S-matrix. For the non-thermal DM scenario, we show the allowed parameter space in terms of the inflaton and DM masses for a given reheating temperature, and compute the comoving free-streaming length to identify the hot, warm and cold DM regimes.}
\keywords{Dark Matter, Inflation, Physics of the Early Universe, Boltzmann Equations, Relic Density} 
\begin{document}
\maketitle

\section{Introduction}\label{sec:intro}
There is overwhelming astrophysical and cosmological evidence for the existence of Dark Matter (DM) in our Universe 
(for a review, see~\cite{Bertone:2004pz}). Assuming the standard $\Lambda$CDM (cosmological constant+Cold Dark Matter) picture of the 
Universe, the recent measurements from the 
Planck mission yield the current matter density in the Universe to be 4.9\% in the form of baryonic matter and 26.6\% as non-baryonic, non-luminous DM, while the remaining 68.5\% is in the form of Dark Energy~\cite{Ade:2013zuv}. Despite all the compelling evidence from its gravitational interaction, the origin and nature of DM are still unknown, and resolving these issues is one of the main goals 
of modern cosmology as well as particle physics. 

On the other hand, cosmological observations such as Planck~\cite{Ade:2013uln} are strongly pointing towards an epoch of  primordial inflation (for a review, see~\cite{Mazumdar:2010sa}), which is considered to be one of the best paradigms to create the seed perturbations for the DM particles to form the observed large-scale structures~\cite{Lyth}. Inflation not only stretches the primordial perturbations on large scales but also dilutes all matter, and 
therefore, it is important that the inflaton must excite the Standard Model 
(SM) degrees of freedom (d.o.f) after the end of inflation for the success of Big Bang 
Nucleosynthesis (BBN)~\cite{BBN}. 

Irrespective of the origin of the inflaton field, whose potential leads to inflation, the inflaton could decay into the SM d.o.f, and also directly to the 
DM particles. The process of creating the entropy happens after the end of inflation during reheating or preheating (for reviews, see~\cite{Allahverdi:2010xz, Mazumdar:2010sa}). If the inflaton is a SM gauge-singlet field $\phi$, it can in principle couple and decay to DM particles $\chi$ with some unknown branching ratio.\footnote{Note that the DM particles can also be created from the scatterings of the inflaton quanta as it happens in the case of preheating~\cite{Kofman:1997yn}, but here we will not discuss this scenario as the detailed computation of such processes requires both analytic and lattice simulations, and a precise definition of the reheat temperature of the Universe which goes well 
beyond the scope of the present paper.}  
As a concrete example, one can envisage a visible-sector inflation scenario within the 
Minimal Supersymmetric Standard Model (MSSM), where inflation can be driven by the superpartners of quarks and leptons~\cite{Allahverdi:2006iq,Allahverdi:2006cx,Allahverdi:2006we}.  Thus, if the DM can have a significant coupling to the inflaton or the moduli field, it can be created rather efficiently from the direct decay of the inflaton or the moduli~\cite{Lyth:1995ka, Moroi:1999zb}.\footnote{During inflation, there could be many fields dynamically present~\cite{Liddle:1998jc} -- some may assist during inflation, and some may obtain quantum-induced vacuum fluctuations to be displaced at very large vacuum expectation values, commonly known as `moduli' in string theory~\cite{de Carlos:1993jw}. They typically couple very weakly via Planck-suppressed interactions. Their decay products could also create DM, see e.g.,~\cite{Kohri:2009ka, Evans:2013nka}.} The initial abundance of such DM particles could be large enough to overclose the Universe, unless their interaction rate is sufficiently larger than the Hubble expansion rate in the early Universe, i.e. $\Gamma_\chi\gg H(t)$, to make them annihilate efficiently into the SM d.o.f. The requirement of not to overproduce the DM poses a stringent constraint on its mass and interaction rates. In fact, a small branching fraction of the inflaton energy density to DM particles 
can be sufficient to overclose the Universe~\cite{Chialva:2012rq}. 


In this paper, we seek a model-independent way to analyze the thermal and non-thermal 
properties of the DM directly produced from the inflaton decay, in terms of their masses, the initial inflaton branching ratio and the strength of DM coupling to the thermal bath. 
For this purpose, we will make the following minimal assumptions: 

\begin{enumerate}
\item The inflaton decays into the SM d.o.f and the DM in a perturbative  scenario; hence on kinematic grounds, $m_\chi < m_\phi/2$. We do not consider non-perturbative DM production processes during the coherent oscillations of the inflaton, e.g., superheavy DM with $m_\chi\gg m_\phi$ for large enough amplitude of the inflaton field~\cite{Kofman:1997yn}, or  fragmentation of the inflaton condensate associated with a global symmetry~\cite{Enqvist:2002rj,Enqvist:2002si}.

 \item The process of reheating is  instantaneous, and the SM d.o.f produced in the inflaton decay quickly thermalize to achieve a local thermodynamic equilibrium (LTE), i.e., both chemical and kinetic equilibrium. Thus we can define a unique reheat temperature $T_R$~\cite{Mazumdar:2013gya} at which the Universe is dominated by relativistic species. Since we assume the DM to be part of the decay products of the inflaton, its initial number density ($n_\chi$) will be  determined in terms of the number density of the inflaton field ($n_\phi$) and the branching ratio of the inflaton decay to DM ($B_\chi$) which should be small in order to have the standard radiation-domination epoch immediately after reheating. 
 \end{enumerate}

We should note here that the analysis presented in this paper based on the simplified picture mentioned above may not be completely valid if the reheating process is {\it not} instantaneous and a significant amount of DM is produced {\it during} reheating through inelastic scatterings between high energy inflaton decay products and the thermal plasma. However, a proper treatment of this issue must include the details of the thermalization process which involve some model-dependent subtleties. In particular, it is important to know {\it when} the inflaton decay products thermalize during the inflaton oscillations. This depends on how the DM interacts with the ambient thermal plasma, such as the SM d.o.f. These thermal interactions lead to some finite temperature effects on DM production rate which will be model-dependent and has to be studied separately. In addition, there exist certain physical situations where the thermalization of the inflaton decay 
products can happen very late, even beyond the complete 
decay of the inflaton. This issue has to be dealt again separately, 
because the DM can still be created before 
the epoch of thermalization, but instead of thermal corrections, there will be finite momentum effects due to the hard-hard, hard-soft and soft-soft scatterings between the inflaton decay products, all of which must be accounted for in a field-theoretically consistent manner. 
Besides these issues, there could also be an epoch of preheating during which the 
DM can be created abundantly, but again this has to be studied in a model-dependent setup. 
Our main goal in this paper is to study the DM abundance from the inflaton decay in a 
{\it model-independent} perspective, and in this regard, we make the simple assumption 
that the thermalization has happened instantly right at the time
of reheating. A more exhaustive analysis of DM creation from the inflaton decay, addressing all the subtleties mentioned above, will be presented elsewhere.

With the assumptions made above, the evolution of the DM particle number density, can be completely described in a model-independent way by its thermally averaged interaction rate $\langle \sigma v\rangle$ with the thermal bath. Depending on the size of $\langle \sigma v\rangle$,  we consider the following three possible scenarios: 
\footnote{Again a concrete example is MSSM inflation in which case the lightest supersymmetric particle, e.g., gravitino or neutralino, could be excited directly from the inflaton decay or its decay products~\cite{Allahverdi:2011aj}, besides the SM d.o.f.  Since gravitinos mostly interact via Planck-suppressed interactions, their abundance will freeze out soon after their production and will be mainly determined by the reheat temperature, while neutralinos have weak interactions and can be quickly brought into kinetic equilibrium (though not necessarily chemical equilibrium) with the bath. Therefore irrespective of how the neutralinos were initially created, their final abundance is always set by the thermal decoupling temperature, as long as $T_R\geq m_\chi$~\cite{Giudice:2000ex}. On the other other, for low reheating temperatures below the standard freeze-out temperature $T_F\sim m_\chi/20$, neutralinos could be a non-thermal DM candidate~\cite{Fornengo:2002db, Gelmini:2006pw}.}

 \begin{enumerate}
 
\item{For large enough $\langle \sigma v\rangle$, the DM particles will quickly reach LTE with the bath, thus losing their initial abundance, and follow the equilibrium distribution until their reaction rate eventually drops below the Hubble rate, after which they will freeze out as a `thermal relic' with a constant comoving number density. This is the standard WIMP scenario~\cite{kolb} in which the final relic abundance is independent of the initial conditions or the details of the production mechanism. Depending on their mass and interaction rate, they could freeze out as a cold, warm, or hot relic~\cite{kolb}. It is well-known that $\langle \sigma v\rangle \sim 10^{-26}~{\rm cm}^3{\rm s}^{-1}$  naturally gives the observed cold DM relic density~\cite{Ade:2013zuv}, almost independent of the DM mass.} 

\item{If the interaction of DM particles with the thermal bath is too small to bring them into full LTE, they will decouple from the bath soon after being produced. Hence, if they are produced abundantly, the final number density will remain large, thus leading to overclosure of the Universe. However, if their initial abundance is negligibly small, the interactions with the bath, although feeble, could still produce some DM particles whose final abundance freezes in at some point as the 
interaction rate eventually becomes smaller than the expansion rate. 
This is the FIMP (Feebly Interacting Massive Particle or Frozen-In Massive Particle) scenario~\cite{Hall:2009bx}.}

 \item For extremely small annihilation rates, the DM particles are never in thermal contact with the bath, and are practically produced decoupled in an out-of-equilibrium condition, and remain non-thermal throughout their evolution. This leads to a super-WIMP (SWIMP)-like scenario~\cite{Feng:2003xh}, where the final abundance is primarily determined by the initial conditions which, in our case, are set by the inflaton mass, 
reheat temperature and branching ratio~\cite{Allahverdi:2002nb, Allahverdi:2002pu}.\footnote{An alternative example where the DM could still be produced from the thermal bath, while its relic abundance is fixed by the reheating temperature is the Non-equilibrium thermal DM scenario~\cite{Mambrini:2013iaa}.} Note that it is also possible to have a non-thermal DM with chemical equilibrium, provided the reheat temperature is smaller than the usual freeze-out temperature so that the DM decouples {\it during} reheating~\cite{Gelmini:2006pw}. We do not consider this case here since we have assumed instant reheating.    

\end{enumerate}

For each of the above three scenarios, we study the evolution of the DM number density by 
numerically solving the Boltzmann equation, and obtain their final relic abundance as a function of their mass and interaction rate for cold, warm as well as hot relics.  We highlight a generic fact that inflaton decay to DM inevitably leads to an overclosure of the Universe for a large range of parameter space, and provides a generic constraint for models of DM with an arbitrary coupling to the inflaton field. 
For a given reheat temperature and initial abundance, we 
show the overclosure region as a function of the 
DM mass and annihilation rate  in a model-independent way. For the thermal WIMP case, we show the complementary constraints on the $(m_\chi,\langle \sigma v\rangle)$ parameter space by taking into account various theoretical as well as experimental 
limits from unitarity, dark radiation, indirect detection, BBN, Cosmic Microwave Background (CMB) and Planck. For the non-thermal production of DM from inflaton decay, we show that a large fraction of the $(m_\phi,~m_\chi)$ parameter space leads to an overclosure for a generic class of hidden sector models of inflation. This an important result in pinning 
down the nature of DM from particle physics point of view and on the allowed region of the inflaton-DM coupling and the branching ratio.
%
%

The rest of the paper is organized as follows: In Section~\ref{sec:Boltzmann}, we briefly review the evolution of DM as governed by the Boltzmann equation. In Section~\ref{sec:prod}, we discuss the production of DM from inflaton decay:  thermal production (both freeze-out and freeze-in scenarios) in Section~\ref{sec:thermal}, and non-thermal production in Section~\ref{sec:non-thermal}. 
In section~\ref{sec:bounds}, we discuss various experimental/observational constraints on DM. 
In Section~\ref{sec:result}, we present our numerical results for both thermal and non-thermal scenarios. 
Our conclusions are given in Section~\ref{sec:conclusion}.
\section{Evolution of DM: a Brief Review}\label{sec:Boltzmann}
The microscopic evolution of the number density $n_\chi$ for any species $\chi$, and the departure from its thermal equilibrium value $n_{\chi,{\rm eq}}$, can be computed exactly by solving the Boltzmann equation~\cite{kolb}
\begin{equation}\label{eq:Boltzmann1}
\frac{dn_\chi}{dt}+3Hn_\chi=-\langle\sigma v \rangle \left( n_\chi^2 - n_{\chi,\text{eq}}^2 \right), 
\end{equation}
%
%
where $\langle \sigma v\rangle$ is the thermally averaged total annihilation rate, $\sigma$ being the total (unpolarized) annihilation cross section, and $v$ being the relativistic relative velocity between the two annihilating particles.\footnote{For the non-relativistic case, $v$ is approximated by the relative velocity $v_r=|{\mathbf v}_1-{\mathbf v}_2|$, while in the general case, it is usually taken to be the M{\o}ller velocity $\bar{v}=\sqrt{({\mathbf v}_1-{\mathbf v}_2)^2-({\mathbf v}_1\times {\mathbf v}_2)^2}$. Here we use the  manifestly Lorentz-invariant definition: $v=\bar{v}/(1-{\mathbf v}_1\cdot {\mathbf v}_2)=\sqrt{(p_1\cdot p_2)^2-m_1^2m_2^2}/(p_1\cdot p_2)$, where $p_1,p_2$ are the four-momenta of the annihilating particles with mass $m_1$ and $m_2$ respectively~\cite{Cannoni:2013bza}.} In the absence of Bose-Einstein condensation or Fermi degeneracy, one can neglect the quantum statistical factors, and write the number density as 
\begin{equation}\label{eq:neq}
n_{\chi}=g_\chi \int \frac{d^3 {\mathbf p}}{(2 \pi)^3}~\exp{\left[-\left(\sqrt{|{\mathbf p}|^2 + m_\chi^2}-\mu_\chi\right)/T\right]}\,,
\end{equation}
where $g_\chi$ is the number of internal (e.g., spin or color) degrees of freedom of $\chi$, $T$ is the temperature, and $\mu_\chi$ is the chemical potential of species $\chi$ (energy associated with change in particle number) which we assume to be zero for the equilibrium number density $n_{\chi ,\text{eq}}$. 
It is useful to express Eq.~(\ref{eq:Boltzmann1}) in terms of the dimensionless quantities $Y_\chi=n_\chi/s$ and $Y_{\chi,{\rm eq}}=n_{\chi,{\rm eq}}/s$ to scale out the redshift effect due to the expansion of the Universe.  
Here,  
\begin{eqnarray}
s = \frac{2 \pi^2}{45} g_{s} T^3 \label{entropy}
\end{eqnarray}
 is the entropy density and $g_{s}$ is the effective number of relativistic degrees of freedom contributing to the total entropy density. Recall that in the early Universe with 
radiation domination, 
$g_s$ is same as the relativistic degrees of freedom $g_\rho$ contributing to the energy density, and also appearing in the Hubble expansion rate: 
\begin{eqnarray}
H(T)=\left(\frac{4\pi^3}{45}\right)^{1/2}g_\rho^{1/2}\frac{T^2}{m_{\rm Pl}}, \label{hubble}
\end{eqnarray}
where $m_{\rm Pl}=1.22\times 10^{19}$ GeV is the Planck mass. Henceforth, we will not distinguish the two, and will take $g_\rho=g_s\equiv g$ which is valid for most of the thermal history of the Universe.\footnote{$g_s$ and $g_\rho$ differ only when there are relativistic species not in equilibrium with photons which happens in the SM for temperatures below the electron mass when the neutrinos have 
already decoupled from the thermal bath, and $e^\pm$ pair-annihilation transfers entropy only to the photons, thus making $g_s$ slightly higher than $g_\rho$ today.} Assuming an adiabatic and isentropic (constant entropy per comoving volume) 
expansion of the Universe, Eq.~(\ref{eq:Boltzmann1}) can be rewritten as~\cite{Srednicki:1988ce, gondolo}
\begin{equation}\label{eq:Y}
	\frac{dY_{\chi }}{dx}= -\frac{s\langle\sigma v \rangle}{Hx}\left(1+\frac{1}{3}\frac{d\ln{g}}{d\ln{T}}\right)\left(Y_{\chi }^2-Y_{\chi ,\text{eq}}^2\right)\,,
\end{equation}
with the introduction of a new independent variable $x= m_\chi/T$.
The current number density $Y_\chi(x_0)$ of the species $ \chi $ is obtained by 
integrating Eq.~(\ref{eq:Y}) from $x=0$ to $x=x_0\equiv m_\chi/T_0$, where $T_0=2.7255(6)$ K is the present 
temperature of the CMB photons~\cite{Fixsen:2009ug}. Knowing $Y_\chi(x_0)$, 
we can compute the relic density of $\chi$, conventionally defined as the ratio of its current mass density,  
$\rho_\chi(x_0)=m_\chi s_0 Y_\chi(x_0)$, and the critical density of the Universe, $\rho_c=3H_0^2/8\pi$. Using the current values for the entropy density $s_0=2889.2~{\rm cm}^{-3}(T_0/2.725~{\rm K})^3$, and the critical mass density $\rho_c= 1.05375(13)\times {10}^{-5} h^2~{\rm GeV}{\rm cm}^{-3}$~\cite{pdg}, we obtain
\begin{equation}\label{eq:omega1}
\Omega_{\chi} h^{2}= 2.74 \times 10^8~ Y_{\chi}(x_0)\left(\frac{m_\chi}{1~{\rm GeV}}\right).
\end{equation}

Eq.~(\ref{eq:Y}) is a form of the Riccati equation for which there is no general, closed-form analytic 
solution. Therefore, the current density $Y_\chi(x_0)$ in Eq.~(\ref{eq:omega1}) has to 
be obtained either by numerically solving Eq.~(\ref{eq:Y}) or by approximating it with an analytic solution in some special cases. 
In the standard $\Lambda$CDM cosmology, the thermal relics decouple from the thermal plasma 
in the radiation-dominated era after inflation, and the decoupling occurs at some freeze-out temperature $T_F$ 
when the annihilation rate $\Gamma_\chi=n_\chi\langle \sigma v\rangle$ drops below the Hubble expansion rate $H$. Depending on the exact value of $x_F=m_\chi/T_F$, one can have the following three scenarios:
%
%
\subsection{Non-relativistic Case} \label{nonrel}
For $x_F\gsim 3$, the DM particles are mostly non-relativistic when they decouple from the thermal plasma. This leads to the usual cold DM scenario with free streaming lengths of sub-pc scale~\cite{Bertone:2004pz}, 
as favored by the standard theory of large-scale structure formation~\cite{Peebles, Primack:1997av}. 
Analytic approximate formulas for their relic abundance have been 
derived in the non-relativistic limit $x_F\gg 1$~\cite{gondolo, Scherrer:1985zt, Kamionkowski:1990ni, Griest:1990kh, Steigman:2012nb}. 
The key point is that the actual abundance $Y_{\chi}$ tracks the equilibrium abundance $Y_{\chi,{\rm eq}}$ during early stages of evolution (for $x\lsim x_*$), 
while at late stages ($x\gsim x_*$),  $Y_{\chi,{\rm eq}}$ is exponentially suppressed and has essentially no 
effect on the final abundance $Y_{\chi}(x_0)$. Here $x_*$ is some intermediate matching point ({\it not} the freeze-out point $x_F$, as commonly assumed) where the deviation from equilibrium starts to grow exponentially. After solving for $x_*$ iteratively as a function of $m_\chi$, $\langle\sigma v\rangle$ and $g_*$ 
(the relativistic degrees of freedom at $x=x_*$), Eq.~(\ref{eq:Y}) can be integrated from $x=x_*$ to $x=x_F$ dropping the $Y_{\chi,{\rm eq}}$ term, to finally obtain an improved  analytic solution for the relic density (in the $s$-wave limit)~\cite{Steigman:2012nb}:
\begin{eqnarray}
\Omega_\chi h^2 = \frac{9.92\times 10^{-28}~{\rm cm}^3{\rm s}^{-1}}{\langle \sigma v\rangle}\frac{x_*}{g_*^{1/2}}\frac{(\Gamma_\chi/H)_*}{1+\alpha_*(\Gamma_\chi/H)_*}\,, \label{eq:omgcold}
\end{eqnarray}   
where the subscript $*$ means the values evaluated at $x=x_*$, and 
\begin{eqnarray}
\alpha_* = \int_{T_F}^{T_*} \frac{dT}{T_*}\left(\frac{g}{g_*}\right)^{1/2}\left(1+\frac{1}{3}\frac{d\ln g}{d\ln T}\right)\,.
\end{eqnarray}
The analytic result in Eq.~(\ref{eq:omgcold}) agrees with the exact numerical result within $\sim 3\%$, almost independent of the DM mass. Note that 
for an arbitrary $l$-wave annihilation, the above formalism can be repeated by Taylor-expanding $\langle \sigma v\rangle$ in powers of $v_r^2\sim 1/x$.  
%
%
\subsection{Relativistic Case}\label{rel} 
In the other extreme limit, where the freeze-out occurs when the $\chi$ particles are still 
relativistic ($x_F\ll 1$), their current relic abundance $Y_{\chi}(x_0)$ is approximated by the equilibrium 
abundance at freeze-out $Y_{\chi,{\rm eq}}(x_F)$~\cite{kolb, Cowsik:1972gh}. In this case, Eq.~(\ref{eq:neq}) gives $n_{\chi,{\rm eq}}=(\zeta(3)/\pi^2)g_{\rm eff}T^3$, where $g_{\rm eff}= g_\chi~(3g_\chi/4)$ for bosonic (fermionic) $\chi$, and $\zeta(x)$ is the Riemann zeta function. Using the entropy density in the relativistic limit as 
given by Eq.~(\ref{entropy}), we obtain $Y_{\chi,{\rm eq}}(x_F)=0.28 g_{\rm eff}/g (x_F)$ which is insensitive to the details of freeze-out. From Eq.~(\ref{eq:omega1}), the present relic density is then given by
\begin{eqnarray}
\Omega_\chi h^2 = 7.62\times 10^{-2}\frac{g_{\rm eff}}{g(x_F)}\left(\frac{m_\chi}{1~{\rm eV}}\right)\,.
\label{eq:omgrel}
\end{eqnarray}  
Relativistic DM particles in our Universe 
will lead to large damping scales $\gsim 10$ Mpc (roughly the size of typical 
galaxy clusters), thereby suppressing the growth of small-scale structures. 
They would predict a top-down hierarchy in the 
structure formation~\cite{Bond:1980ha, Bond:1983hb}, with small structures forming by fragmentation of larger ones, while observations have shown no convincing evidence of such effects, thereby imposing stringent upper limits on these `hot' DM species. For instance, the SM neutrino contribution to the non-baryonic DM relic density is currently constrained to be $\Omega_\nu h^2\leq 0.0062$ at 95\% confidence level (CL)~\cite{pdg}. Thus, hot DM cannot yield the total observed DM density in 
our Universe~\cite{Ade:2013zuv}, and if it exists,\footnote{Recently, the presence of a hot DM component at $3\sigma$ CL has been proposed to resolve the inconsistencies of the Planck measurements with other observations, such as the 
current Hubble rate, the galaxy shear power spectrum and galaxy cluster counts~\cite{Hamann:2013iba}.} must coexist with other cold/warm components. For example scenarios of such multi-component DM, see~\cite{Chialva:2012rq, Zurek:2008qg, Feldman:2010wy, Winslow:2010nk, Dienes:2011ja}. 
%
%
%
\subsection{Semi-relativistic Case}\label{semrel}
In the intermediate regime $x_F\sim 1$, the $\chi$ particles are semi-relativistic when they 
decouple from the thermal bath. The improved analytic treatment of~\cite{Steigman:2012nb}, as  
discussed in Section~\ref{nonrel}, is not applicable in this case, since the thermally averaged cross section $\langle \sigma v\rangle$ involves multiple integrals, and 
cannot be expanded in a Taylor series of the velocity-squared. One way is to approximate the cross section by interpolating between its relativistic and non-relativistic expressions. Following this approach, it was shown~\cite{Drees:2009bi} that the Maxwell-Boltzmann distribution can still be used to compute $\langle \sigma v\rangle$, and the more appropriate Fermi-Dirac or Bose-Einstein distributions are only needed for the calculation of the freeze-out abundance $Y_{\chi,{\rm eq}}(x_F)$. Note that although the current observations do not rule out the possibility of the whole DM density being comprised of warm DM species (see e.g.,~\cite{Lin:2000qq, Hisano:2000dz, Gelmini:2006vn}), there exist strong constraints from observations of early structure, 
in particular from Lyman-$\alpha$ forest data~\cite{Boyarsky:2008xj, Viel:2013fqw}.    

On the other hand, if the interaction of the DM particles with the thermal bath is not large enough, they may not come into full LTE before they decouple from the plasma. In such cases, their current relic density $Y_\chi(x_0)$ in 
Eq.~(\ref{eq:omega1}) will also depend on the initial abundance, and hence, on the production mechanism. This is discussed in the following section with a simple production mechanism.   
%
%
%
\section{DM from Inflaton  Decay} \label{sec:prod} 
As discussed in Section~\ref{sec:intro}, we assume that the DM particles $\chi$ directly couple to the inflaton field $\phi$ so that it can be produced in the perturbative inflaton decay for $m_\chi<m_\phi/2$.\footnote{This is the necessary condition required solely due to kinematic reasons, and could be sufficient, for instance, for fermionic DM 
coupling to the inflaton through a $\phi\bar{\chi}\chi$ term in the Lagrangian. For more complicated inflaton decay chains involving many particles, a more stringent kinematic condition may be required. Also we do not consider derivative couplings of the inflaton with DM, which could be the case for axion DM, for example.} The initial energy density stored in the inflaton field is $\rho_\phi\approx n_\phi m_\phi$ which is transferred to the decay products at the end 
of inflation, thereby (re)heating the Universe with a temperature $T_R$. Assuming the Universe to be 
radiation dominated immediately after inflation, the total energy density is given by $\rho_r=(\pi^2/30)gT_R^4$.\footnote{Note that the assumption $\rho_\phi=\rho_r=(\pi^2/30)gT_R^4$ which determines the reheat temperature may not be correct for all possible inflaton or moduli coupling to the 
SM d.o.f. This definition is correct for large inflaton coupling to matter, i.e. $\alpha_\phi \geq 10^{-7}$~\cite{Mazumdar:2013gya}. Typically the moduli coupling to the SM d.o.f and DM will be very small, i.e. $\alpha_\phi \sim (m_\phi/m_{\rm pl})^2$. Therefore, a more rigorous treatment of the reheating scenario is required for the case of moduli decay.} 
Hence, the initial DM number density is given by
\begin{equation}\label{eq:nchi,i}
n_{\chi,{\rm in}} = B_\chi n_{\phi} \simeq B_\chi\frac{\pi^2 g}{30}\frac{T_R^4}{m_\phi}, 
\end{equation}
where $B_\chi$ is the branching ratio of the inflaton decay to DM. Since we are interested in model-independent constraints on the DM parameter space, we keep our discussion general in terms of the branching ratio, without specifying its exact formula in terms of the DM-inflaton couplings, their masses, and the $n$-body decay kinematics (for $n\geq 2$, depending on the specific DM candidates).  

Once produced, depending on the strength of their interaction with the thermal plasma, they could either thermalize fully/partially with the thermal bath or could remain non-thermal throughout their evolution. In the former case, their current relic density will be determined by their freeze-out  
abundance, independent of the initial abundance set by the inflation parameters.  This is also true for the freeze-in scenario, provided the initial abundance is small compared to the thermal abundance. 
In the non-thermal case, however, their final number density is essentially the same as their initial abundance, only redshifted by the Hubble expansion rate. These two different scenarios are discussed below in somewhat details, 
with some numerical examples.    
%
%
\subsection{Thermal DM}\label{sec:thermal}
%
%
In this case, depending on their thermal annihilation rate $\langle \sigma v\rangle$, the $\chi$ particles can either reach full LTE (i.e. both kinetic and chemical equilibrium) with the plasma before decoupling or decouple from the plasma before the full equilibrium could be established. The former case occurs for large annihilation rates, 
which enable the $\chi$ particles to attain equilibrium soon after their production. In this case, 
the $\chi$ particles follow the equilibrium distribution until they freeze out at a certain stage, depending on the exact value of the interaction rate. Thus, the initial abundance is irrelevant for their final relic density. 
In the latter case, the annihilation rates are not large enough to bring the $\chi$ particles into full LTE, and hence, their final abundance is determined by the annihilation rate as well as the initial abundance given by Eq.~(\ref{eq:nchi,i}). For given inflaton and DM masses, 
the final relic abundance $\Omega_\chi h^2$ 
will exceed the observed value for a large reheat temperature and/or large 
branching ratio of the inflaton to DM, thus overclosing the Universe. If the initial abundance is small, the DM particles can still be produced from the thermal plasma unless the interaction rate is utterly negligible. The dominant production in this case occurs at temperatures $T\gsim m_\chi$ when the interaction rate is still larger than the Hubble rate, and as the interaction rate drops below the Hubble rate, the relic abundance will freeze-in. We discuss below both freeze-out and freeze-in scenarios for the DM produced from inflaton decay, and give a numerical example for each case to illustrate the magnitudes of the interaction cross section, as compared to the well-known thermal WIMP scenario. 
%
%
\subsubsection{Freeze-out}\label{sec:fo}
In this case, the final relic abundance of the DM species is set by the freeze-out abundance which is determined by the freeze-out temperature. This is obtained by solving the Boltzmann equation (\ref{eq:Y}) for $Y_\chi$. 
To calculate the freeze-out abundance more precisely, we track the evolution of the quantity $\Delta_\chi=(Y_\chi-Y_{\chi,{\rm eq}})/Y_{\chi,{\rm eq}}$ which represents the departure from equilibrium.  From Eq.~(\ref{eq:Y}), 
the evolution equation for $\Delta$ is obtained to be of the form
\begin{eqnarray}
\frac{d\ln{(1+\Delta)}}{d\ln x} = -\frac{d\ln{Y_{\chi,{\rm eq}}}}{d\ln x}-\frac{\Gamma_{\chi,{\rm eq}}}{H}\left(1+\frac{1}{3}\frac{d\ln g}{d\ln T}\right)\frac{\Delta(2+\Delta)}{1+\Delta}, \label{eq:delta}
\end{eqnarray}
where $\Gamma_{\chi, {\rm eq}}=n_{\chi, {\rm eq}}\langle \sigma v\rangle=Y_{\chi, {\rm eq}}s\langle \sigma v\rangle$ is the equilibrium annihilation rate, and $H(x)$ can be readily obtained from Eq.~(\ref{hubble}).  
%
%
For a Maxwell-Boltzmann distribution, the equilibrium number density and thermally averaged annihilation cross section are respectively given by~\cite{gondolo}  
\begin{eqnarray}
Y_{\chi,{\rm eq}}(x) &=& \frac{45}{4\pi^4}\frac{g_\chi}{g}x^2K_2(x), \label{eq:Ygon} \\
\langle \sigma v\rangle (x) &=& \frac{1}{8m_\chi^4 T K_2^2(x)}\int_{4m_\chi^2}^\infty d\bar {s}(\bar{s}-4m_\chi^2)\sqrt{\bar{s}} K_1\left(\frac{\sqrt {\bar{s}}}{T}\right)\sigma(\bar{s}),  \label{eq:siggon}
\end{eqnarray}
where $K_n(x)$ is the $n$-th order modified Bessel functions of the second kind, and 
$\sqrt{\bar{s}}$ is the center-of-mass energy. Strictly speaking, Eq.~(\ref{eq:siggon}) is only applicable for the non-relativistic case with $x\gg 1$. However, as noted in~\cite{Scherrer:1985zt, Drees:2009bi}, this is a good approximation (within 3\% accuracy) even for the semi-relativistic case with $x\sim 1$. For the relativistic case $x\ll 1$, the final abundance is simply the equilibrium abundance, as given in Eq.~(\ref{eq:omgrel}).  

Following the strategy developed in~\cite{Steigman:2012nb} to solve Eq.~(\ref{eq:delta}) for $\Delta$, we note that in the early stages of evolution, $Y_\chi$ tracks $Y_{\chi,{\rm eq}}$ closely, and hence, $\Delta, d\Delta/dx \ll 1$. In this case, the left-hand side of Eq.~(\ref{eq:delta}) can be safely dropped, thus leading to 
\begin{eqnarray}
\frac{\Delta(2+\Delta)}{1+\Delta} = -\frac{d\ln{Y_{\chi,{\rm eq}}}}{d\ln x}\frac{H}{\Gamma_{\chi,{\rm eq}}}\left(1+\frac{1}{3}\frac{d\ln g}{d\ln T}\right)^{-1}. \label{eq:delta2}
\end{eqnarray}
As the $\chi$ particles start freezing out with increasing $x$, $\Delta$ increases exponentially, eventually becoming much larger than 1. Thus for some intermediate value of $x=x_*$, $\Delta\sim {\mathcal O}(1)$, and for $x>x_*$, it grows exponentially. We define $x_*$ when $\Delta(x_*)\equiv \Delta_* = 1/2$,\footnote{As verified in~\cite{Steigman:2012nb}, other alternative choices of $\Delta_*$ change the final result only by about 0.1\%.} and solve Eq.~(\ref{eq:delta2}) iteratively for $x_*$ as a function of $m_\chi$, $\langle \sigma v\rangle$ and $g_*$. For the logarithmic derivative of of $g(T)$, we use the calculations of~\cite{Laine} for the SM relativistic d.o.f. 
For the cases with no phase transition 
around $T_*=m_\chi/x_*$, $g(T)$ is almost constant, and hence, this term can be ignored in Eq.~(\ref{eq:delta2}). Once the value of $x_*$ is found, we can determine $T_*=m_\chi/x_*$ and $Y_{\chi}(x_*)=(3/2)Y_{\chi,{\rm eq}}(x_*)$ (corresponding to $\Delta_*=1/2$). The actual freeze-out temperature $T_F$ is somewhere below $T_*$, since at $T=T_*$, $(\Gamma_\chi/H)_*$ is still larger than 1~\cite{Steigman:2012nb}. 

For $x>x_*$, $Y_\chi\gg Y_{\chi,{\rm eq}}$, and hence, the $Y^2_{\chi,{\rm eq}}$ term in Eq.~(\ref{eq:Y}) can be dropped. Integrating from $x=x_*$ to $x=x_0$, we obtain the present relic abundance: 
\begin{eqnarray}
Y_{\chi}(x_0) = \left[\frac{1}{Y_{\chi}(x_*)}+\int_{x_*}^{x_0}dx\frac{s\langle \sigma v\rangle}{Hx}\left(1+\frac{1}{3}\frac{d\ln g}{d\ln T}\right)\right]^{-1}\,,
\label{eq:ychi}
\end{eqnarray} 
which can be used in Eq.~(\ref{eq:omega1}) to compute $\Omega_\chi h^2$.    
To perform the integration in Eq.~(\ref{eq:ychi}), we need to know the $x$-dependence of $\langle \sigma v\rangle$ using Eq.~(\ref{eq:siggon}) which is one of the key quantities that determine the current relic density. In general, one can find an ansatz for $\langle \sigma v\rangle$ which smoothly interpolates between the non-relativistic and relativistic regimes. 
For simplicity, we will use the ansatz for an $s$-wave annihilation of two Dirac fermions~\cite{Drees:2009bi}: 
\begin{equation}
\langle\sigma v \rangle = \frac{\alpha_\chi^2 m_\chi^2}{16 \pi} ~ \left(  \frac{12}{x^2}+\frac{5+4 x}{1+x}\right), \label{drees}
\end{equation}
where $\alpha_\chi$ denotes the coupling constant of the four-fermion interaction, which we will treat as a free parameter. This approach works well for DM species that freeze out between $ 0.5 \lesssim x_F \lesssim 15 $. For $ x_F > 15 $ (roughly corresponding to $m_\chi > 10$ MeV), the particles are already non-relativistic at 
decoupling, and hence, one can expand $\langle\sigma v \rangle $ in a Taylor series 
in terms of the averaged relative velocity:
\begin{equation}
\langle\sigma v \rangle = a + b \left\langle v_r^2\right\rangle  + {\mathcal O}\left(\left\langle v_r^4\right\rangle\right) = a + \frac{b'}{x} + {\mathcal O}\left( \frac{1}{x^2} \right)\,.
\end{equation}
For $s$-wave annihilation, only the first term is considered, and in this case, Eq.~(\ref{eq:ychi}) simplifies further to finally yield the relic density given by Eq.~(\ref{eq:omgcold}). We note that this approximation of using a constant value for $\langle\sigma v\rangle$ also works well in the semi-relativistic case, and induces an error of only about 6\%, as compared to using the ansatz given by Eq.~(\ref{drees}).    

From Eq.~(\ref{eq:ychi}), it is clear that the final abundance is {\it inversely} proportional to the thermal annihilation rate. Thus, the larger the cross section, the longer the DM particles stay in equilibrium with the thermal bath, and hence, the lower the final abundance. This is true for both cold and warm DM cases, while for the hot DM case, the freeze-out is insensitive to the interaction cross section, as discussed in Section~\ref{rel}.

\begin{figure}[t!]
\centering
\includegraphics[width=9cm]{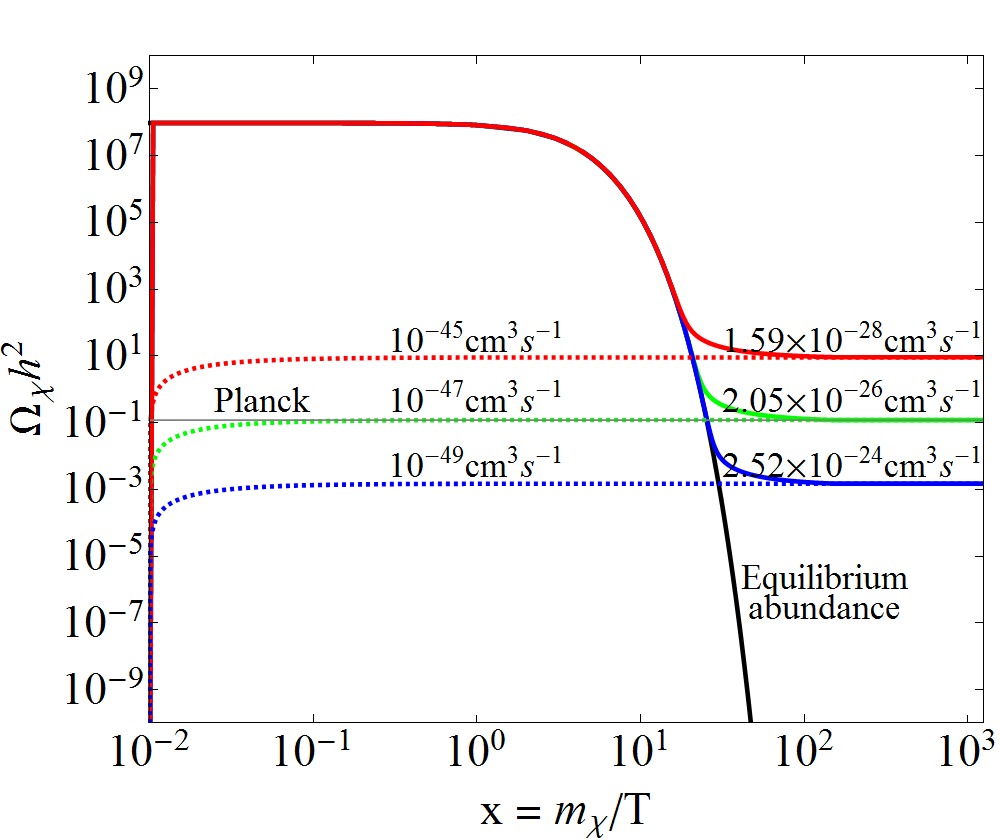}
\caption{The illustration of freeze-out and freeze-in scenarios in the evolution of thermal DM abundance as a function of $x=m_\chi/T$ for different annihilation rates. Here we have chosen $m_\chi=100$ GeV and for the initial conditions, $m_\phi=10^{13}$ GeV, $T_R=10$ TeV, $B_\chi=10^{-15}$. The horizontal band gives the observed relic density from Planck data~\cite{Ade:2013zuv}.}
\label{fig:thermal}
\end{figure}

The dependence of the current relic abundance on the annihilation rate for the thermal DM which has frozen out is illustrated in Figure~\ref{fig:thermal}. Here we have chosen $m_\chi=100$ GeV. The solid black line shows the equilibrium distribution which is constant in the extreme relativistic regime ($x\ll 3$), and exponentially suppressed in the non-relativistic regime ($x\gg 3$), as can also be seen from Eq.~(\ref{eq:Ygon}) by taking the asymptotic limits of the Bessel function. For large enough annihilation rates, the DM particles quickly thermalize, thereafter following the equilibrium evolution until their freeze-out, and the final relic abundance is independent of the initial abundance. 
The observed relic density as measured by Planck, shown as the horizontal band, is obtained for the thermal annihilation rate of $\langle \sigma v\rangle = 2\times 10^{-26}~{\rm cm}^3{\rm s}^{-1}$, as shown by the solid red line. As the annihilation rate decreases, the DM freezes out earlier (with smaller $x_F$), thus giving a larger relic density. 

\subsubsection{Freeze-in} \label{fi}   

In this scenario, the DM particles are very weakly coupled to the bath, and hence, cannot 
reach full thermal equilibrium with the bath before decoupling. However, the feeble interactions with the thermal bath (either directly~\cite{Hall:2009bx} or mediated by a portal~\cite{Blennow:2013jba})  could still populate the DM, until the interaction rate drops below the Hubble rate when the DM abundance will freeze in. In this case, the final abundance is {\it directly} proportional to the interaction strength; the larger the interaction cross section is, the more DM particles are produced. In this sense, freeze-in can be viewed as the opposite process to freeze-out. 

The final relic density in the freeze-in scenario will in general be determined by both the interaction cross section and the initial abundance which in turn depends on the reheat temperature and the branching ratio of the inflaton in our case. Note that the decoupling in this case occurs for small values of $x_F$, where the equilibrium abundance $Y_{\chi,{\rm eq}}$ is independent of $x$, as can be seen from Eq.~(\ref{eq:Ygon}): 
\begin{eqnarray}
Y_{\chi,{\rm eq}}(x\ll 1) = \frac{45\zeta(3)}{2\pi^2}\frac{g_{\rm eff}}{g} \label{yeq}
\end{eqnarray}
Also since the DM particles decouple very soon after being produced, the annihilation cross section as well as the number of relativistic degrees of freedom can be treated as constant with respect to $x$ during this short period of time. Hence, the  general Boltzmann equation (\ref{eq:Y})  can be approximated in this case to the following simple form:
\begin{eqnarray}
\frac{dY_\chi}{dx}=\sqrt{\frac{\pi}{45}}g^{1/2}\langle \sigma v\rangle
\frac{m_\chi m_{\rm Pl}}{x^2}(Y_\chi^2-Y_{\chi,{\rm eq}}^2) \simeq -\frac{A}{x^2}(Y_\chi^2-B), \label{boltz11} 
\end{eqnarray}
  where $A$ and $B$ are constants in $x$. Eq.~(\ref{boltz11}) has a simple analytic solution in terms of the initial values $x_i=m_\chi/T_R$ and  $Y_{\chi,{\rm in}}$, where the latter can be obtained from Eqs.~(\ref{eq:nchi,i}) and (\ref{entropy}):
\begin{eqnarray}
Y_{\chi,{\rm in}}=\frac{n_{\chi,{\rm in}}}{s(T_R)} \simeq \frac{3}{4} B_\chi \frac{T_R}{m_\phi}.
\label{ychi-in}
\end{eqnarray}
In the limit $x\to \infty$, the expression for $Y_\chi(x)$ simplifies further, and the final relic density can then be obtained using Eq.~(\ref{eq:omega1}). This has two contributions: 
\begin{eqnarray}
\Omega_\chi h^2 &=&  2.06\times 10^8~ B_\chi \frac{m_\chi}{m_\phi}\left(\frac{T_R}{1~{\rm GeV}}\right)\nonumber\\
&&+g^{1/2}\langle \sigma v\rangle m_{\rm Pl}m_\chi\left(\frac{T_R}{1~{\rm GeV}}\right)\left(5.6\times 10^6 \frac{g^2_{\rm eff}}{g^2}-4.1\times 10^7 B_\chi^2\frac{T_R^2}{m_\phi^2}\right)
\label{omg-fin}
\end{eqnarray}
where the first term represents the non-thermal contribution which {\it only} depends on the initial abundance, and the other two terms represent the thermal contribution which also depend on the interaction 
rate.  Note that the analytic expression (\ref{omg-fin}) is valid as long as $m_\chi\ll T_R$ otherwise the thermal production will be delayed to lower values of temperature (or higher values of $x$) when the equilibrium distribution in Eq.~(\ref{boltz11}) may no longer be flat, but exponentially decaying. For the freeze-in scenario, it is usually assumed that the initial abundance is negligible, so that the final abundance is solely determined by the interaction strength in Eq. (\ref{omg-fin}), as in the freeze-out scenario. This is illustrated in Figure~\ref{fig:thermal} for a typical choice of parameters: $m_\chi=100$ GeV, $m_\phi=10^{13}$ GeV, $T_R=10$ TeV, and $B_\chi=10^{-15}$ so that the initial abundance given by Eq.~(\ref{ychi-in}) is negligible. The different dashed lines in Figure~\ref{fig:thermal} correspond to the freeze-in scenario with various interaction rates, and hence, different final abundances. Note that the final abundance {\it increases} with increasing interaction rate, in contrast with the freeze-out scenario (the solid lines) where the final abundance {\it decreases} with increasing interaction rate. As shown here, the observed relic abundance shown by the gray horizontal band can be obtained in the freeze-in scenario for an interaction rate of $10^{-47}~{\rm cm}^3{\rm s}^{-1}$, which is much smaller than the typical value of 
$2\times 10^{-26}~{\rm cm}^3{\rm s}^{-1}$, as in the freeze-out scenario.   

We should mention here that there could be other thermal production mechanisms for the DM in specific models, depending on its interaction with the SM particles and/or the model construction for the beyond SM sector. For instance, a keV-scale sterile neutrino DM can be produced by the Dodelson-Widrow mechanism~\cite{Dodelson:1993je}, which is very similar to the freeze-in mechanism discussed above.  
\subsection{Non-thermal DM}\label{sec:non-thermal}
\begin{figure}[t!] 
\begin{center} 
		\includegraphics[height=7.7cm]{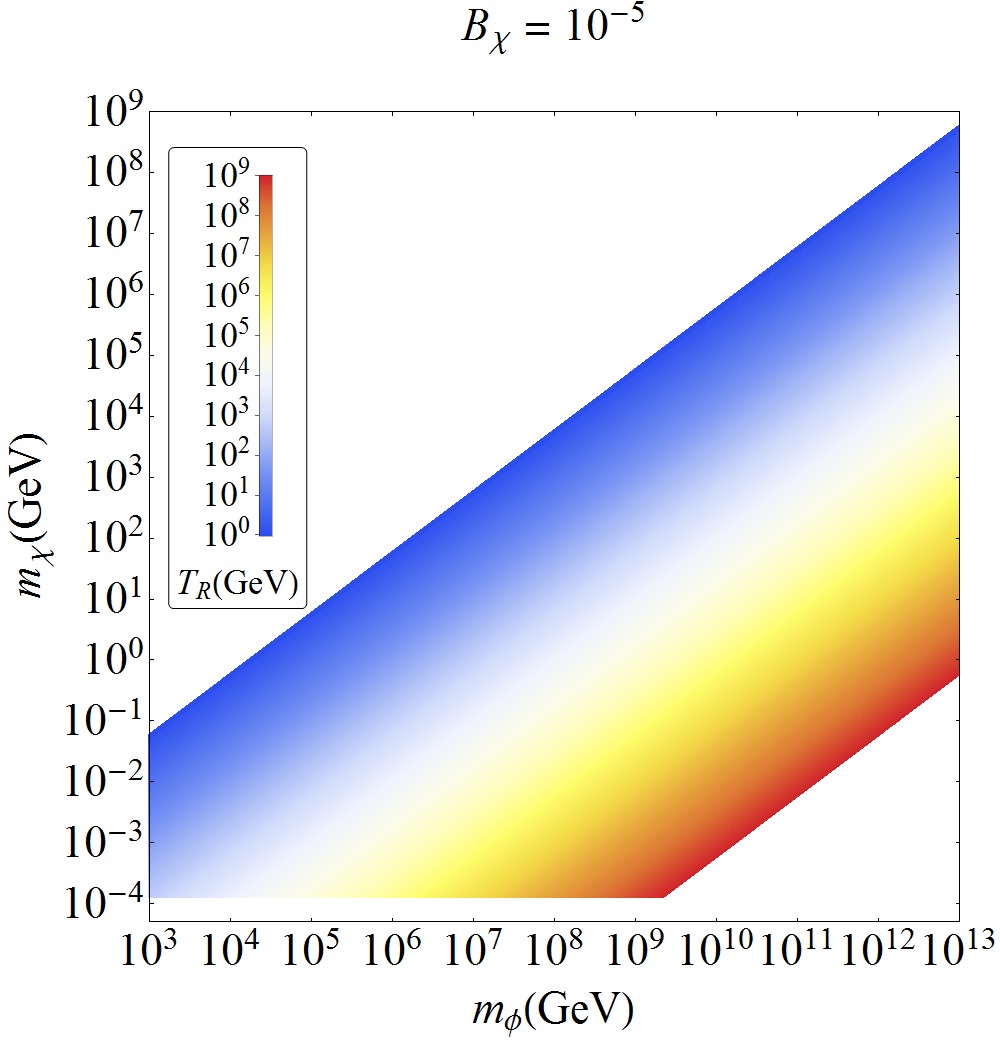}  
		\includegraphics[height=7.7cm]{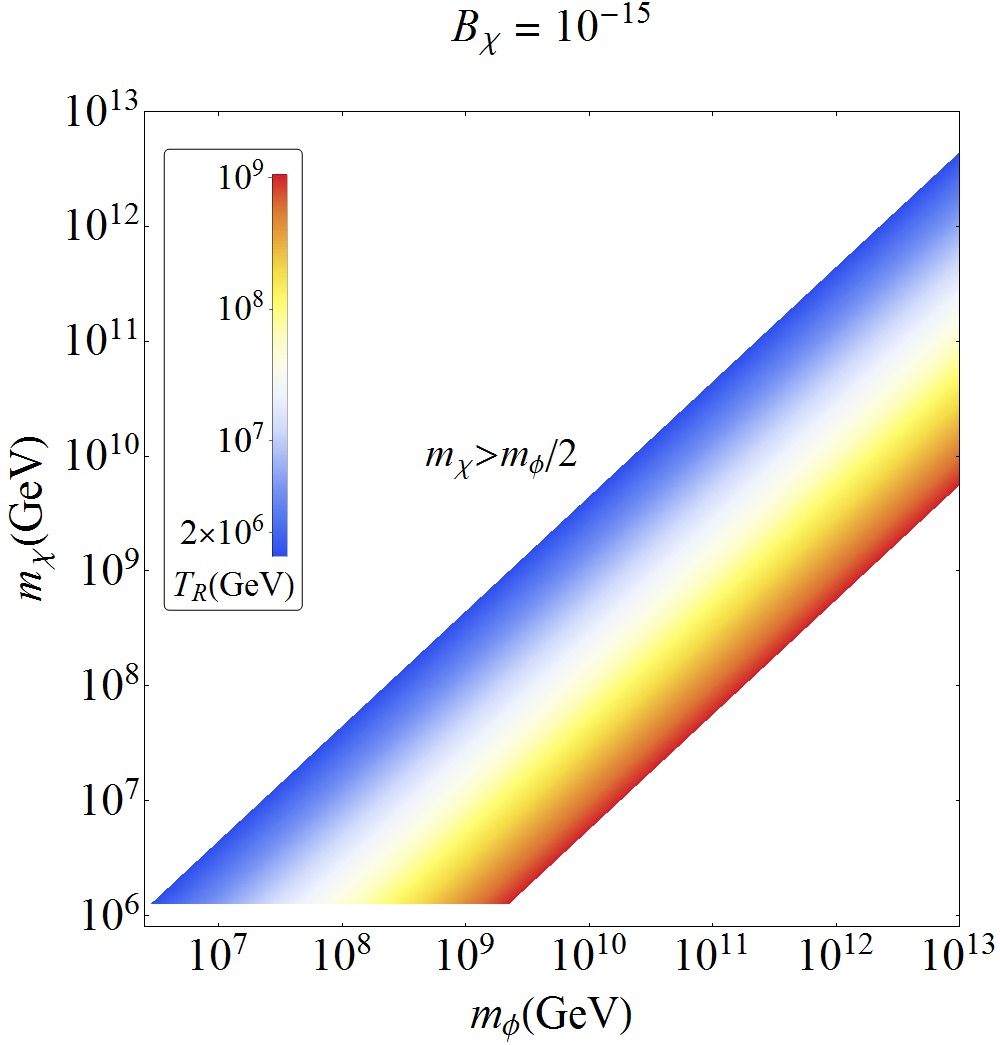}
\caption{The colored contours show the reheat temperature values required to give the correct relic density 
for non-thermal DM as a function of the inflaton and DM masses, for a given inflaton branching ratio.}
\label{fig:ntp} 
\end{center} 
\end{figure}

For very small cross sections, the DM particles are produced already decoupled from the thermal bath, and hence, the thermal production in Eq.~(\ref{omg-fin}) is negligible compared to the initial 
abundance, which could be sizable for large branching ratios. In this case, the annihilation rate, and hence, the right-hand side of Eq.~(\ref{eq:Y}) can be neglected, thus leading to $dY_{\chi}/dx \simeq 0 $. Hence, the final relic 
abundance is completely determined by the initial one given by Eq.~(\ref{ychi-in}). 
Using the general expression (\ref{eq:omega1}), this yields the non-thermal relic DM density
\begin{eqnarray}
\Omega_\chi h^2 \simeq 2.06\times 10^8~ B_\chi \frac{m_\chi}{m_\phi}\left(\frac{T_R}{1~{\rm GeV}}\right), \label{eq:omega4}
\end{eqnarray} 
which can also be identified with the first term on the right-hand side of Eq.~(\ref{omg-fin}). 
Thus for super-weak interaction rates, the final abundance only depends on the reheat temperature and inflaton branching fraction for given DM and inflaton masses.\footnote{Similar results were 
obtained in~\cite{Campos} for superheavy metastable DM candidates. Our result is valid for all non-thermal DM production mechanisms as long as it is a perturbative process.} Some illustrative cases for the non-thermal DM 
are shown in Figure~\ref{fig:ntp} for two typical values of the branching ratio 
$B_\chi=10^{-5}$ and $10^{-15}$. The choice of small values of $B_\chi$ will be justified below. The various contours show the reheat temperature values required to obtain the correct relic density $\Omega_\chi h^2 = 0.12$ for given values of the inflaton and DM masses. These plots were obtained by numerically solving the Boltzmann equation (\ref{eq:Y}) for a typical annihilation rate $\langle \sigma v\rangle =10^{-60}~{\rm cm}^3{\rm s}^{-1}$ (see Section~\ref{sec:result} for details) following the procedure mentioned above, but the results agree quite well with the approximate analytic formula given in Eq.~(\ref{eq:omega4}). From Figure~\ref{fig:ntp} it is clear that as the inflaton branching fraction increases, the allowed range of the DM mass shifts to lower values in order to satisfy the observed relic density, in accordance with Eq.~(\ref{eq:omega4}). We have shown the results for the inflaton mass $m_\phi$ in the range $ 10^3$ - $10^{13}$ GeV, the reheat temperature $T_R$ between 1 - $ 10^{9}$ GeV and for the DM mass $m_\chi\leq m_\phi/2 $. Note that a late-time entropy production would induce various cosmological effects, leading to a lower limit on the reheat temperature of about 1 MeV from BBN constraints~\cite{Kawasaki:1999na, Kawasaki:2000en}, which, when combined with the CMB and large scale structure data, improves to about 4 MeV at 95\% CL~\cite{Hannestad:2004px}. However, for $T_R$ below the QCD phase transition scale, $\Lambda_{\rm QCD}= {\mathcal O}(100)$ MeV, the estimation of the number density of particles from inflaton decay suffers from large QCD uncertainties due to the non-perturbative hadronization effects. Hence, we have used a conservative value, which is one order of magnitude above the hadronization scale, as the minimum value of $T_R$ in Figure~\ref{fig:ntp} (left panel).\footnote{One should also note that for $T_R$ below the electroweak phase transition scale, $\Lambda_{\rm EW} = {\mathcal O}(100)$ GeV, one cannot use the standard electroweak sphaleron processes to explain the observed baryon asymmetry, and must invoke a post-sphaleron baryogenesis mechanism~\cite{Dimopoulos:1987rk, Cline:1990bw, Babu:2006xc, Babu:2008rq, Babu:2013yca}. }

For $m_\chi\ll m_\phi$, the non-thermal DM directly pair-produced from the inflaton decay will have a large velocity at the time of matter-radiation equality ($t_{\rm eq}$), unless the reheat temperature is sufficiently high to make the velocity small due to redshift.  The comoving free-streaming length of a non-thermal DM 
at matter-radiation equality is given by 
\begin{eqnarray}
\lambda_{\rm fs} & = & \int_{t_D}^{t_{\rm eq}}dt \frac{v_\chi(t)}{a(t)}, \label{ntlam}
\end{eqnarray}
where $a(t)$ is the scale factor, $t_D$ is the time at inflaton decay, and 
\begin{eqnarray}
v_\chi(t) = \frac{|{\mathbf p_\chi}|}{E_\chi} \simeq \frac{\frac{m_\phi}{2}\frac{a(t_D)}{a(t)}}{\sqrt{m_\chi^2+\left[\frac{m_\phi}{2}\frac{a(t_D)}{a(t)}\right]^2}}
\end{eqnarray}
is the magnitude of the velocity of the DM particle. Integrating Eq.~(\ref{ntlam}), and requiring that 
$\lambda_{\rm fs}\lsim$ 1 Mpc, from Lyman-$\alpha$ constraints (for warm/cold DM), one obtains a lower limit on the 
reheat temperature~\cite{Takahashi:2007tz} 
\begin{eqnarray}
T_R \gsim 5\times 10^4~{\rm GeV} \left(\frac{g}{200}\right)^{-1/4}\left(\frac{1~{\rm GeV}}{m_\chi}\right)\left(\frac{m_\phi}{10^{12}~{\rm GeV}}\right)
\end{eqnarray}
Combining this with Eq.~(\ref{eq:omega4}), and requiring that $\Omega_\chi h^2\leq 0.12$ to satisfy the observed relic density for cold/warm DM relics, we derive an upper limit on the branching ratio of the inflaton decay to DM: 
$B_\chi \lsim 0.01(g/200)^{1/4}$. This is complementary to what is already expected from the fact that for a  
standard Cosmology, $B_\chi$ must be small in order to have a radiation-domination epoch immediately after reheating, followed by matter domination only at a late stage. 


\section{Experimental Constraints}\label{sec:bounds}
In this section, we summarize the various experimental constraints on the DM properties relevant for our analysis.
\subsection{Overclosure}\label{sec:5.1}
For any DM candidate, we must ensure that it does not lead to an overclosure of the Universe. Thus, we set the upper limit on the relic density of our $\chi$ particles coming from inflaton decay using the 
observed value $\Omega_{\rm CDM}h^2=0.1199\pm 0.0027~~(68\%~{\rm CL; Planck+WP})$~\cite{Ade:2013zuv}, 
which combines the Planck temperature data with WMAP polarization data at low multipoles. We do not set a lower limit  on $\Omega_\chi$ since for the cases in which the $\chi$ particles do not account for the total observed abundance, 
the remaining fraction can be obtained by invoking a hidden-sector/multi-component DM scenario (see e.g.,~\cite{Chialva:2012rq, Zurek:2008qg, Feldman:2010wy, Winslow:2010nk, Dienes:2011ja}). 
\subsection{Unitarity}\label{sec:5.2}
The partial-wave unitarity of the scattering matrix, together with the conservation of total energy and momentum, impose a generic upper bound on the cross section of thermal DM annihilation into the $j$-th partial wave~\cite{Griest}:
\begin{eqnarray}
\sigma_j \leq \frac{4\pi(2j+1)}{m_\chi^2v^2}\left(1-\frac{v_r^2}{4}\right), \label{eq:uni}
\end{eqnarray}
where $v_r=2\sqrt{1-4m_\chi^2/\bar{s}}$ is the relative velocity between the two annihilating particles in the center-of-mass frame with total energy $\sqrt {\bar{s}}$.  Assuming that the $s$-wave piece with $j=0$ dominates in the partial-wave expansion, we obtain an upper bound on the thermally averaged annihilation rate $\langle \sigma v\rangle$ as a function of the DM mass from Eq.~(\ref{eq:siggon}), where $\sigma$ is replaced with $(\sigma_0)_{\rm max}$ from Eq.~(\ref{eq:uni}). Since the current abundance of a non-relativistic thermal relic scales as $\Omega_\chi\propto 1/\langle \sigma v\rangle$, the observed DM relic density constrains the mass of the thermal relic to be $m_\chi\lsim 130$ TeV to satisfy the unitarity bound. Note however that this bound may not be applicable when the higher partial-waves are not suppressed, as is the case when the DM particles decouple from the thermal bath while still being relativistic.   
\subsection{Planck} \label{sec:5.3}
Precision measurements of the CMB angular power spectrum by Planck put stringent constraints on the number of effective neutrino species ($N_{\rm eff}$), which parametrizes the total radiation energy density of the Universe:
\begin{eqnarray}
\rho_r=\rho_\gamma\left[1+\frac{7}{8}\left(\frac{T_\nu}{T_\gamma}\right)^4 N_{\rm eff}\right],\label{rhor}
\end{eqnarray}
where $\rho_\gamma=(\pi^2/15)T^4$ is the energy density of photons, and the neutrino-to-photon temperature ratio $T_\nu/T_\gamma=(4/11)^{1/3}$ assuming exactly three neutrino flavors and their instantaneous decoupling.  In the 
standard cosmological model, $T_\nu/T_\gamma$ is slightly higher than $(4/11)^{1/3}$ due to partial reheating of neutrinos when electron-positron pairs annihilate transferring their entropy to photons, thus giving $N_{\rm eff}=3.046$~\cite{Mangano:2005cc}. Now if the DM species remains in thermal 
equilibrium with the neutrinos or electrons and photons after neutrino decoupling, and transfers its 
entropy to them during its annihilation after it decouples at a later stage, it can increase or decrease the value of $N_{\rm eff}$ as we decrease the DM mass. 
Using the constraints on $N_{\rm eff}$ from Planck~\cite{Ade:2013zuv}, together with the helium abundance $Y_p$,~\cite{Boehm:2013jpa} derived a robust {\it lower} bound of 2-10 MeV on the thermal DM mass, depending on whether it is a fermion (Dirac/Majorana) or scalar (real/complex) and whether it was in equilibrium with neutrinos or with electrons and photons.  

Another generic lower bound on the cold DM mass can be obtained using the CMB and matter power spectrum observations which place an upper bound on the DM temperature-to-mass ratio: $T/m_\chi \leq 1.07 \times 10^{-14}~ (1+z)^2$~\cite{Armendariz-Picon:2013jej}. Evaluating this bound at matter-radiation equality with a redshift of $z_{\rm eq}=3391$~\cite{Ade:2013zuv} and $T_{\gamma,{\rm eq}}\simeq 0.77$ eV~\cite{kolb}, we obtain a lower limit of $m_\chi\gsim 6.5$ keV, which is much weaker than the limit derived in~\cite{Boehm:2013jpa} using $N_{\rm eff}$. 
\subsection{Dark Radiation}\label{sec:5.4}
The Planck constraints on $N_{\rm eff}$ can also be used to set an upper limit on the amount of dark 
radiation at decoupling. From Eq.~(\ref{rhor}), the radiation energy density apart from the photon and SM neutrino contribution is given by 
\begin{eqnarray}
\Omega_{\rm dark} h^2 = \frac{7}{8}\left(\frac{4}{11}\right)^{4/3}\Delta N_{\rm eff}~\Omega_\gamma h^2, \label{omgr}
\end{eqnarray}
where $\Omega_\gamma h^2=2.471\times 10^{-5}(T/2.725)^4$ is the CMB radiation density~\cite{pdg}, and $\Delta N_{\rm eff}=N_{\rm eff}-3.046$. Using the 95\% CL measured value of $N_{\rm eff}=3.30^{+0.54}_{-0.51}$ from Planck+WMAP-9 polarization data+SPT high-multipole measurement+Baryon Acoustic Oscillation measurements from large scale structure surveys (Planck+WP+highL+BAO)~\cite{Ade:2013zuv}, we obtain an upper limit on the amount of dark radiation from Eq.~(\ref{omgr}): $\Omega_{\rm dark} h^2\leq 4.46\times 10^{-6}$. This also sets the upper limit on the relic density of hot DM species. In order to obtain the mass range in which the thermal DM species decouple while being relativistic, we calculate their free-streaming length~\cite{Boyanovsky:2007ba}: 
\begin{equation}\label{eq:lambda_th}
\lambda_{\text{fs}}(z) = (1.4 \times 10^{-2}~{\rm  Mpc})g^{-1/3}
\sqrt{\frac{1+z}{I_0}}\left(\frac{1~{\rm keV}}{m_\chi}\right),
\end{equation}
where $z$ is the redshift, and $I_0=[\int_{0}^{\infty} f^{(0)}(y)dy]/[\int_{0}^{\infty} y^2f^{(0)}(y)dy]$ is a dimensionless ratio, given in terms of the comoving energy distribution function $f(y)=1/\exp{[\sqrt{y^2 + x^2}+1]}$ with $x=m_\chi/T$ and the superscript $(0)$ refers to the current value of the distribution. From the Ly-$\alpha$ constraints, we require $\lambda_{\rm fs}(0)\lsim 1$ Mpc for cold/warm DM candidates. For concreteness, we will impose the dark radiation upper limit from Eq.~(\ref{omgr}) for the parameter space corresponding to $\lambda_{\rm fs}(0)>2$ Mpc.    
\subsection{BBN and CMB}\label{sec:5.5}
The late annihilation of DM particles (after freeze out) can deposit hadronic and/or electromagnetic energy in the primordial plasma, thereby altering the history of nucleosynthesis (BBN)~\cite{Jedamzik:2006xz, Hisano:2009rc, Henning:2012rm} and recombination (CMB)~\cite{Galli:2009zc,Slatyer:2009yq,Huetsi:2009ex,Cirelli:2009bb,Hutsi:2011vx,Galli:2011rz,Cline:2013fm, Madhavacheril:2013cna}. These effects depend only on the type and rate of energy injection into the thermal bath, thus allowing to set rather model-independent  bounds on the annihilation rate, especially for DM masses in the MeV-GeV range. During nucleosynthesis, the injection of hadronic and/or electromagnetic energy can affect the abundance of nuclei via (i) raising the neutron-to-proton ratio and therefore the primordial $^4$He abundance, and (ii) high energy nucleons and photons disassociating nuclei. During recombination, the injected electromagnetic energy ionizes hydrogen atoms, which results in an increased number of free electrons, causing the broadening of the surface of last scattering, and results in scale-dependent changes to the CMB temperature and polarization power spectra, especially in the low multipole modes. The precision measurements of BBN and CMB from WMAP and Planck data have been used to set upper bounds on the DM annihilation cross section $ \langle\sigma v \rangle $, as a function of the DM mass~\cite{Jedamzik:2006xz, Hisano:2009rc, Henning:2012rm, Cirelli:2009bb,Hutsi:2011vx,Galli:2011rz,Cline:2013fm, Madhavacheril:2013cna}. 
\subsection{Indirect Detection} \label{sec:5.6}
After freeze-out, the relic annihilations of WIMP DM may be indirectly observed by searching for their annihilation products such as charged particles, photons and neutrinos (for a review, see~\cite{Cirelli:2012tf}). In fact, a number of indirect detection experiments have observed an excess of electrons and positrons in the charged cosmic ray flux, and this was recently confirmed with the precision measurements by AMS-02~\cite{ams2}.  Assuming a possible DM contribution to this positron excess and using the high quality of AMS-02 data,~\cite{Bergstrom:2013jra} has performed a spectral analysis to put stringent constraints on the DM annihilation cross section for various leptonic final states.\footnote{A DM interpretation of the AMS-02 positron excess is still viable, if the DM annihilates to four-lepton final states~\cite{DM1, DM3, DM4, hooper0}.} Similar constraints were obtained for the DM annihilation into hadronic final states~\cite{Evoli:2011id, Cirelli:2013hv} in order to explain the absence of a corresponding excess in the cosmic-ray antiproton flux in the PAMELA data~\cite{Adriani:2008zq, Adriani:2010rc}.

The DM annihilation to various SM final states can also lead to an observable photon flux which can be produced either by direct DM annihilation (`prompt' gamma-rays) or by inverse Compton scattering and synchrotron emission of the electrons and positrons created in the DM annihilation. These photon signals are preferentially searched for in regions with high DM densities and/or regions with reduced astrophysical background. The Fermi-LAT, with its unprecedented sensitivity to gamma rays in the MeV-TeV energy range, has performed deep searches for line spectrum (mono-energetic gamma-rays due to direct DM annihilation)~\cite{Fermi-LAT:2013uma} as well as continuum spectrum  (through DM annihilation into intermediate states)~\cite{Ackermann:2012qk}.\footnote{There exists yet another class of spectral signature, namely, box-shaped gamma-ray spectrum, which arises if the DM annihilates/decays into intermediate particles which further decay into photons~\cite{Ibarra:2012dw}. The cross-section limits derived using this feature are currently comparable to those obtained using the line-like spectral feature.} They have derived additional constraints on the DM annihilation cross section from the isotropic diffuse 
gamma-ray emission in the galactic halo~\cite{Ackermann:2012rg}, nearby galaxy clusters~\cite{Ackermann:2010rg}, and nearby dwarf spheroidal galaxies~\cite{Ackermann:2013yva}. Similar constraints were also derived from the galactic center region for various DM density profiles~\cite{Hooper:2012sr}. Complementing the Fermi-LAT range toward higher energies, the HESS collaboration has performed a number of DM searches up to multi-TeV DM masses~\cite{Abramowski:2011hc, Abramowski:2011hh, Abramowski:2013ax}. 
  
The DM annihilation can also produce neutrinos which, like gamma-rays, can travel essentially unabsorbed through the galaxy, and can be observed at large neutrino detectors on Earth. Constraints on the DM annihilation rate were derived by the IceCube experiment from the upper limits on the high-energy neutrino fluxes from the galactic halo~\cite{Abbasi:2011eq}, galactic center~\cite{Abbasi:2012ws}, dwarf galaxies and clusters of galaxies~\cite{Aartsen:2013dxa}. These limits are currently somewhat weaker than the gamma-ray limits for low DM masses, but become competitive at larger DM masses. Combining the Fermi-LAT data on the diffuse gamma-ray and the IceCube data on diffuse neutrino flux, robust constraints were derived on the DM 
annihilation rate for heavy DM masses (1 TeV - $10^{10}$ GeV)~\cite{Murase:2012xs}.        

\section{Results and Discussion}\label{sec:result}\label{sec:6}
\begin{figure}[t!]
\centering
\includegraphics[width=15cm]{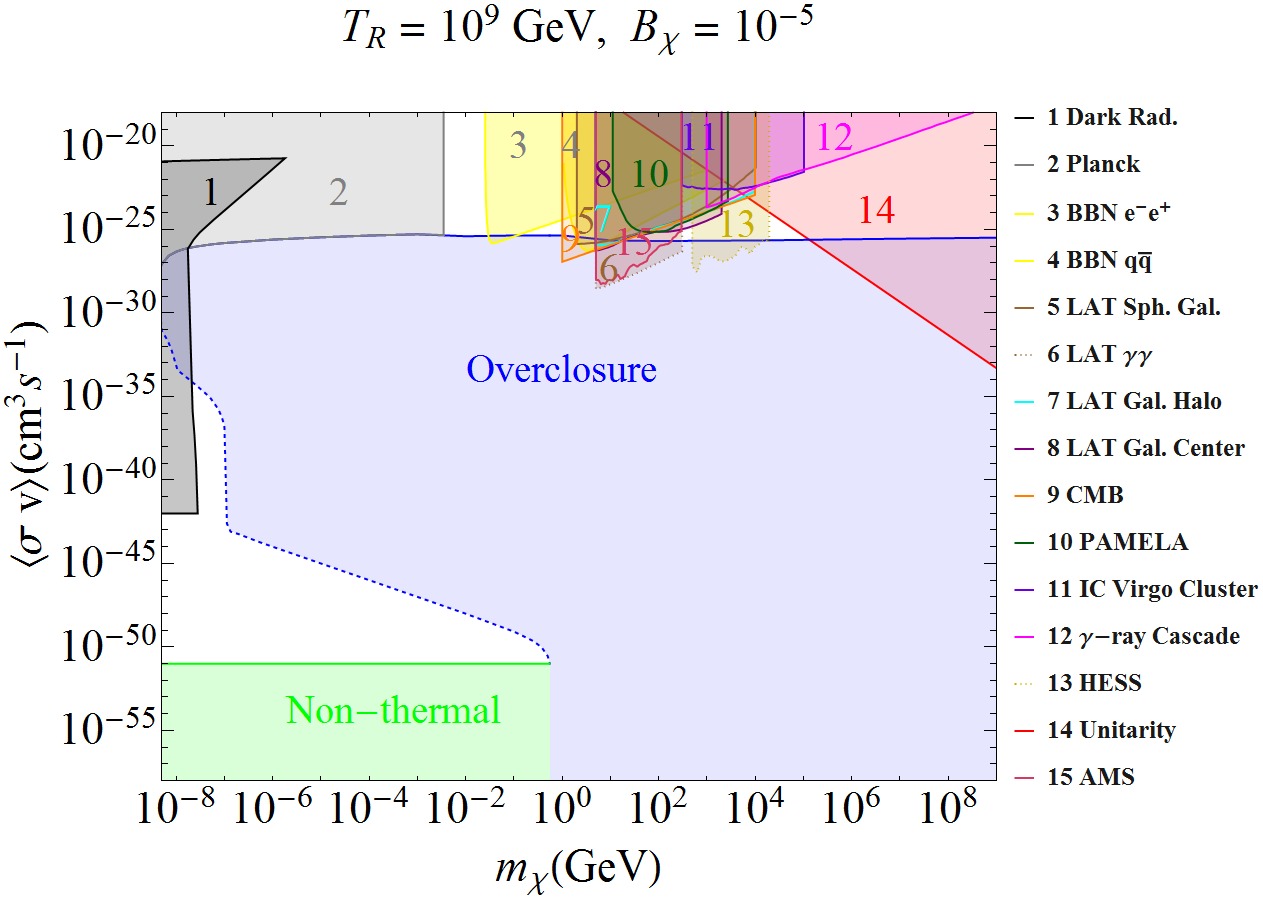}
\caption{Model-independent constraints on the DM annihilation rate as a function of the DM mass for both thermal and non-thermal production mechanisms. Here we have chosen $m_\phi=10^{13}$ GeV, $T_R=10^9$ 
GeV and $B_\chi=10^{-5}$ as the initial parameters for the DM evolution. The blue-shaded region is excluded from relic density constraints, and the observed relic density is obtained at its boundary (shown by the solid and dotted blue 
lines). The green-shaded region at the bottom represents the non-thermal DM scenario, while in the rest of the parameter space, the DM can be fully/partly thermalized with the primordial bath. The various colored-shaded regions in the thermal region are excluded (under certain assumptions) by the constraints given in Section~\ref{sec:bounds}; see text for details.}\label{fig:scan1}
\end{figure}

\begin{figure}[t!]
\centering
\includegraphics[width=15cm]{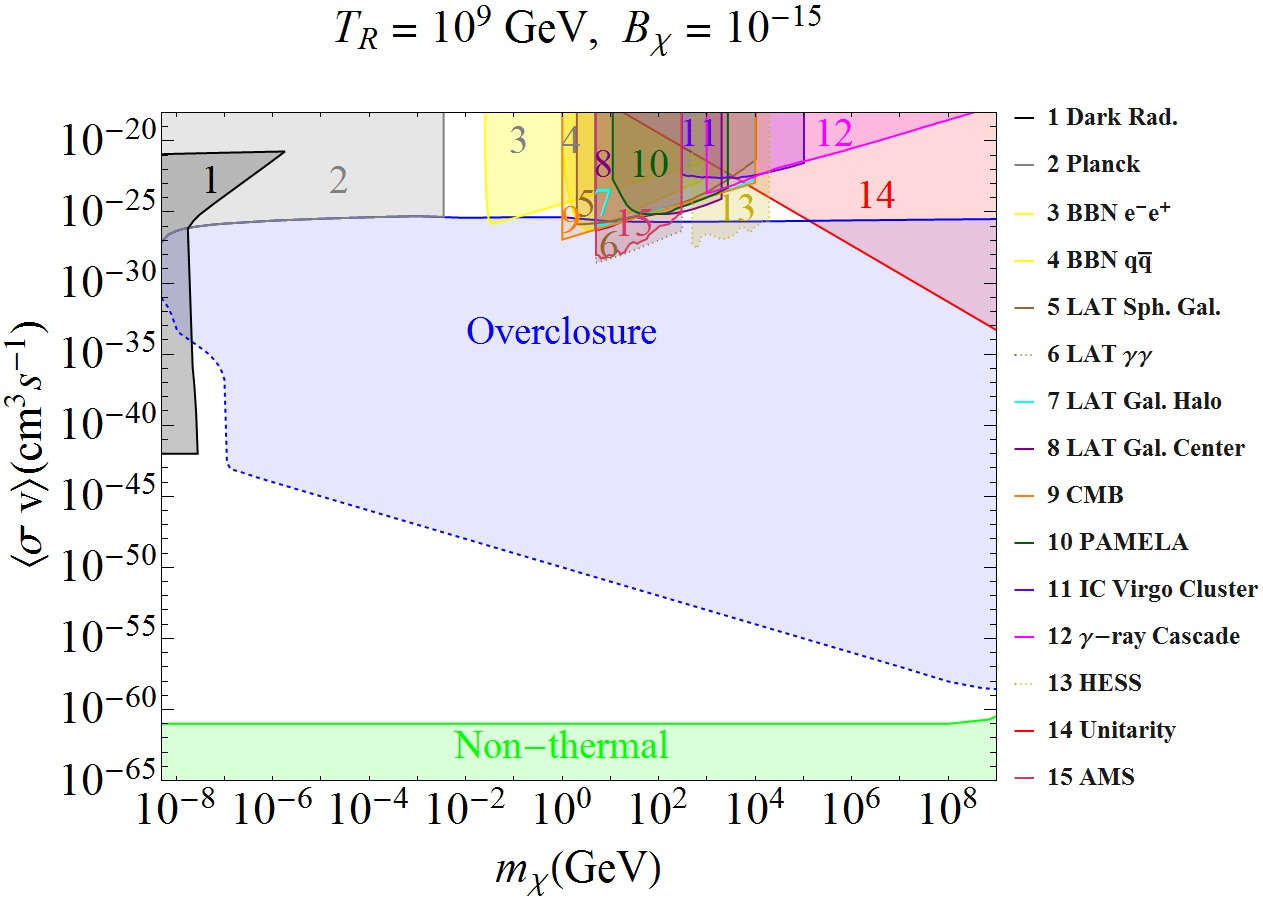}
\caption{The parameters and labels are the same as in Figure~\ref{fig:scan1}, except for 
$B_\chi=10^{-15}$ to show the dependence on the initial conditions.} 
\label{fig:scan1b}
\end{figure}

\begin{figure}[t!]
\centering
\includegraphics[width=15cm]{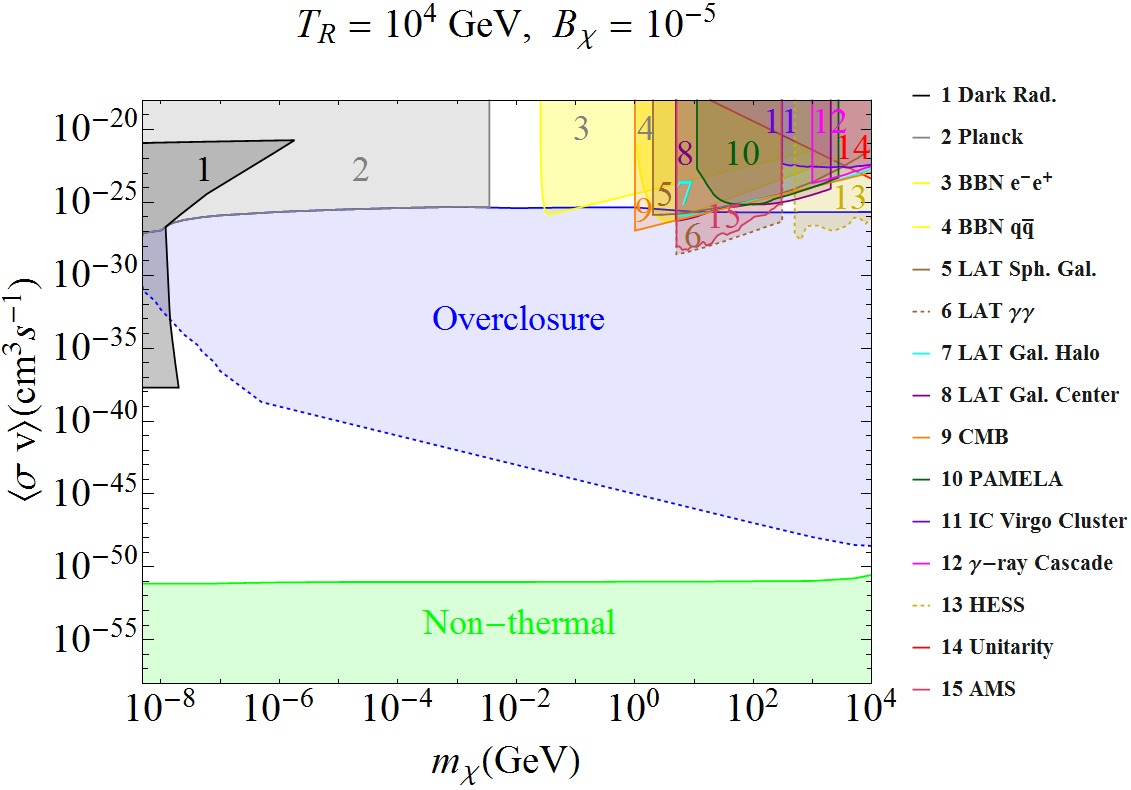}
\caption{Same as Figure~\ref{fig:scan1}, except for $T_R=10^4$ GeV.}\label{fig:scan2}
\end{figure}

\begin{figure}[h!]
\centering
\includegraphics[width=15cm]{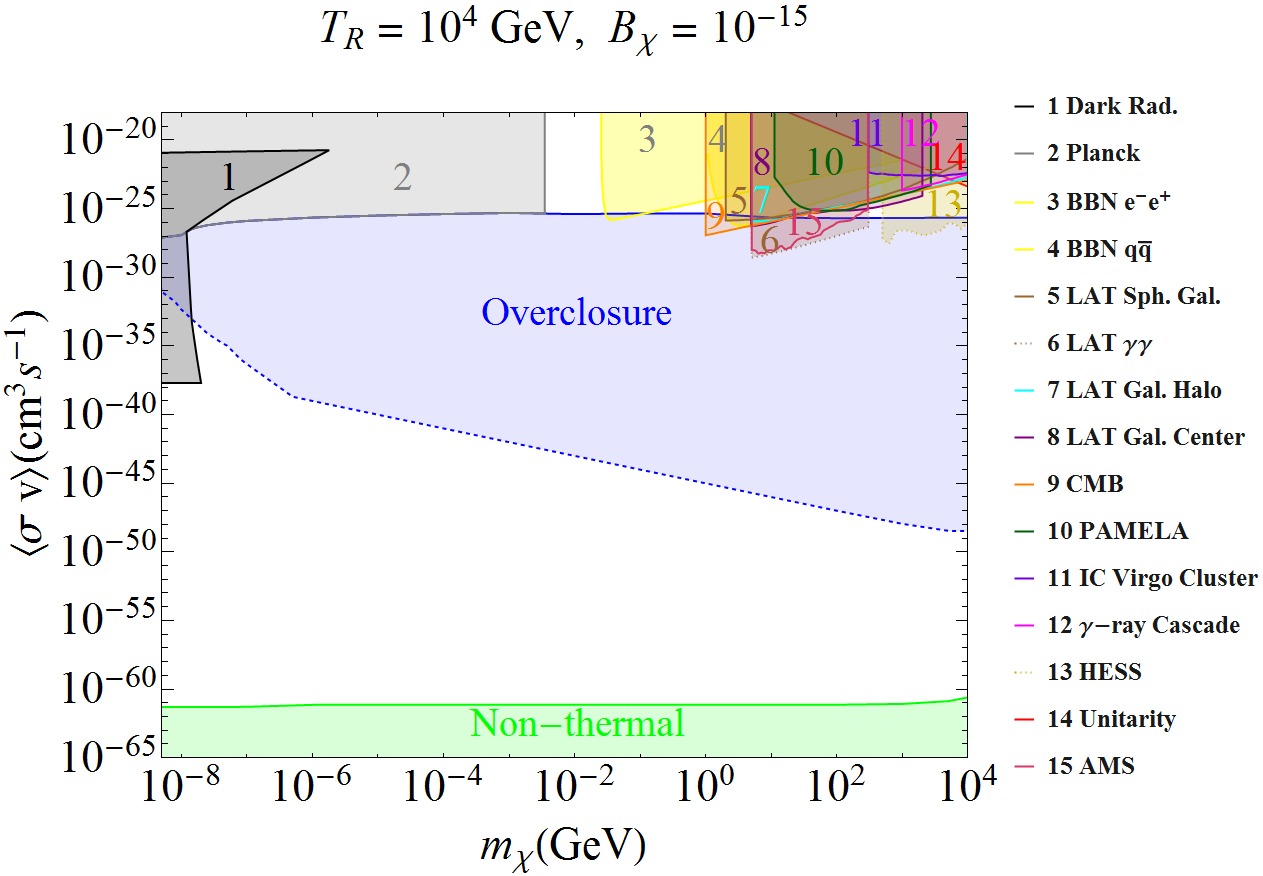}
\caption{Same as Figure~\ref{fig:scan1}, except for $T_R=10^4$ GeV and $B_\chi=10^{-15}$.}\label{fig:scan2b}
\end{figure}

Using the model-independent approach outlined in Section~\ref{sec:prod}, we solve the Boltzmann equation (\ref{eq:Y}) numerically for the evolution of DM produced from inflaton decay. Here we assume an $s$-wave annihilation, and take the annihilation rate $\langle \sigma v\rangle$ to be a free parameter.\footnote{For a $p$-wave annihilation,  $\langle \sigma v\rangle$ depends on the temperature, and hence, cannot be taken as a free parameter in the Boltzmann equation. Our assumption also obliterates additional complications that could arise in special cases such as co-annihilation and resonant annihilation. However, these are highly model-dependent effects, and we cannot easily generalize our results to such scenarios. A more accurate, model-specific numerical analysis for the relic density can be done with publicly available codes~\cite{Gondolo:2004sc,Belanger:2006is,Backovic:2013dpa}. }
Both thermal and non-thermal regions are identified in the ($ m_\chi $, $ \langle\sigma v \rangle $) parameter space. Our results are shown in Figures~\ref{fig:scan1}-\ref{fig:scan2b} for a fixed inflaton mass $m_\phi=10^{13}$ GeV. We consider two typical values of the reheat temperature $T_R=10^9$ GeV and $10^4$ GeV, and branching ratios of the inflaton decay $B_\chi=10^{-5}$ and $10^{-15}$ for our illustration purposes. We have considered the DM masses only below the reheating temperature, and do not analyze scenarios in which DM could be produced {\it during} preheating or 
reheating (e.g., the WIMPzilla scenario~\cite{Chung:1998rq}).

For each case shown in Figures~\ref{fig:scan1}-\ref{fig:scan2b}, we calculate the current relic density of the thermal DM to show the overclosure region (blue-shaded) which rules out a wide range of the parameter space, irrespective of the initial choice of parameters. In the remaining allowed parameter space, we identify the region with very low annihilation rates belonging to the non-thermal DM scenario (green-shaded region) 
since for 
such extremely small interaction rates, the DM particles cannot achieve LTE, and decouple soon after being produced essentially with their initial abundance, as discussed in Section~\ref{sec:non-thermal}. So the overclosure condition for non-thermal DM will be determined solely by the initial conditions, and this will be discussed in Section~\ref{sec:6.2}.   
\subsection{Thermal Case} \label{sec:6.1}
In the thermal DM regime, the region above the overclosure region with large annihilation rate belongs to the freeze-out scenario, while in the white region below the overclosure one with small annihilation rate belongs to either freeze-out or freeze-in scenario, depending on the initial conditions. The observed value of the relic density is obtained at the boundary between these regions with the overclosure region (shown by the solid and dotted blue lines). The thermal freeze-out region with large annihilation rate is severely constrained by many experimental searches, as discussed in Section~\ref{sec:bounds}, some of which are shown by the shaded regions 1-15 in Figures~\ref{fig:scan1}-\ref{fig:scan2b}, and also summarized below: 
\begin{itemize}
\item Region 1 is excluded by the dark radiation constraint, as discussed in Section~\ref{sec:5.4}. In this region, the comoving free-streaming length is greater than 2 Mpc, thus corresponding to a hot DM regime, while the relic density $\Omega_\chi h^2$ exceeds the upper limit of $4.46\times 10^{-6}$ derived from the Planck data~\cite{Ade:2013zuv}. 

\item Region 2 is excluded by the recent Planck measurements of the effective number of neutrino species, as discussed in Section~\ref{sec:5.3}, assuming that the DM interacts with neutrinos or electrons and 
photons after the neutrino decoupling. This sets a robust lower bound of order of MeV on the thermal DM mass with large interaction rates. The precise value of the lower bound depends on whether the DM is a scalar or fermion and on whether it couples to neutrinos or to electrons and photons. The bound shown by region 2 assumes a Majorana fermion DM coupling to neutrinos. Note that~\cite{Boehm:2013jpa} had originally derived this limit for a cold DM candidate, but this is generically applicable as long as the interaction rate is large enough to keep the DM in LTE after the 
neutrino decoupling, thus transferring entropy at a late stage and affecting $N_{\rm eff}$.   

\item Regions 3 and 4 are excluded by the BBN data, as discussed in Section~\ref{sec:5.5}, and assuming DM annihilation into electron-positron and quark-antiquark pairs respectively~\cite{Henning:2012rm}. Similarly, the region 9 is excluded by constraints derived from a combination of the CMB power spectrum measurements from Planck, WMAP9, ACT and SPT, and low-redshift datasets from BAO, HST and supernovae~\cite{Madhavacheril:2013cna}.

\item Region 5 is excluded by the Fermi-LAT limit at 95\% CL derived using the diffuse gamma-ray flux from a combined analysis of 15 dwarf spheroidal galaxies, for an NFW DM density profile and assuming DM annihilation into tau-antitau final states~\cite{Ackermann:2013yva}. Region 7 is excluded by the $3\sigma$ Fermi-LAT limit obtained using the diffuse gamma-ray emission in the Milky Way halo, assuming an NFW DM distribution and for annihilation into bottom-antibottom quark pairs~\cite{Ackermann:2012rg}. Region 8 is excluded by a similar analysis using the Fermi-LAT data from galactic center~\cite{Hooper:2012sr}. The corresponding limits for other SM final states are weaker, and are not shown 
here for clarity purposes.  

\item Region 6 is excluded by the Fermi-LAT 95\% CL upper limit on the cross section of DM annihilation to two photons from a dedicated search for the gamma-ray line spectrum~\cite{Fermi-LAT:2013uma}. Region 13 is excluded from a complementary search for the line 
spectrum  by HESS~\cite{Abramowski:2013ax}. Note that these limits, although very stringent, can be evaded in most of the popular WIMP DM models, since the direct annihilation to photon final states is suppressed due to loop effects.

\item Region 10 is excluded by the measurements of the antiproton flux from PAMELA, and assuming the DM annihilation 
to $b\bar{b}$  final states~\cite{Cirelli:2013hv}. These limits are applicable only for hadronic final states. Similarly, region 15 is excluded by the 95\% CL upper limits, derived from the AMS-02 data, on the DM annihilation cross section for $e^+e^-$ final state~\cite{Bergstrom:2013jra}. The corresponding limits for other leptonic final states are somewhat weaker, and hence, are not shown here. 

\item Region 11 is excluded by the IceCube upper limit on the DM annihilation cross section for neutrino final states for the Virgo galaxy cluster including subhalos~\cite{Aartsen:2013dxa}. The corresponding limits for other final states as well as from searches in galactic halo~\cite{Abbasi:2011eq} and galactic center~\cite{Abbasi:2012ws} are somewhat weaker.

\item Region 12 is excluded by the cascade gamma-ray constraints obtained using the Fermi-LAT diffuse gamma-ray background data up to very high energies~\cite{Murase:2012xs}. The corresponding limits derived using the IceCube high-energy neutrino data are stronger at higher DM masses, but weaker than the unitarity constraint 
(see Section~\ref{sec:5.2}).  

\item Region 14 is excluded by the unitarity constraints~\cite{Griest}, as discussed in Section~\ref{sec:5.2}, which sets an upper limit on the CDM mass of about 130 TeV for the allowed region, and rules out heavy thermal DM, even 
with annihilation rates many orders of magnitude below the thermal annihilation rate. This theoretical constraint is the most stringent one for very heavy DM, and is applicable as long as the DM is produced thermally.    
\end{itemize}
Note that for the indirect detection constraints, we have shown only a few of them (typically the most stringent ones) for illustration purposes. Most of these limits have limited applicability, as they were derived assuming DM annihilation into a particular final state, and could be evaded in specific models where some of these annihilation channels might be suppressed due to various reasons. Also note that additional constraints on the annihilation cross section for a given DM mass might be derived using possible correlations with the DM direct detection 
cross section limits~\cite{Akerib:2013tjd} and collider search limits from mono-jet~\cite{Chatrchyan:2012me, ATLAS:2012ky} and mono-photon~\cite{Chatrchyan:2012tea, Aad:2012fw} final states with large missing energy.  In the absence of a collider signal for DM, model-independent constraints can be derived on the mass and interactions of a generic WIMP DM candidate from direct and indirect detection searches~\cite{Beltran:2008xg}.   

The other allowed thermal DM parameter space, namely, the region with very low interaction rates such as the 
FIMP scenario, is hard to constrain from the existing experimental limits. Various experimental tests of the freeze-in mechanism by measurements at colliders or by cosmological observations were outlined in~\cite{Hall:2009bx}. However, these signals depend very much on the particular freeze-in scenario under consideration, and hence, it is difficult to derive model-independent constraints in the $(m_\chi,\langle \sigma v\rangle)$ parameter space, except for the 
generic dark radiation constraint as shown in Figures~\ref{fig:scan1}-\ref{fig:scan2b}. Just to give an example of 
additional model-specific bounds, a keV-scale sterile neutrino DM, which has a small interaction rate due to its mixing with the active neutrinos, could radiatively decay to an active neutrino and a photon which will lead to a mono-energetic X-ray line~\cite{Boyarsky:2009ix}, the absence of which puts severe constraints on such keV-scale DM  models, including their production mechanisms~\cite{Merle:2013gea}.   
\subsection{Non-thermal Case}\label{sec:6.2}
\begin{figure}[t!]
\centering
\includegraphics[width=7.6cm]{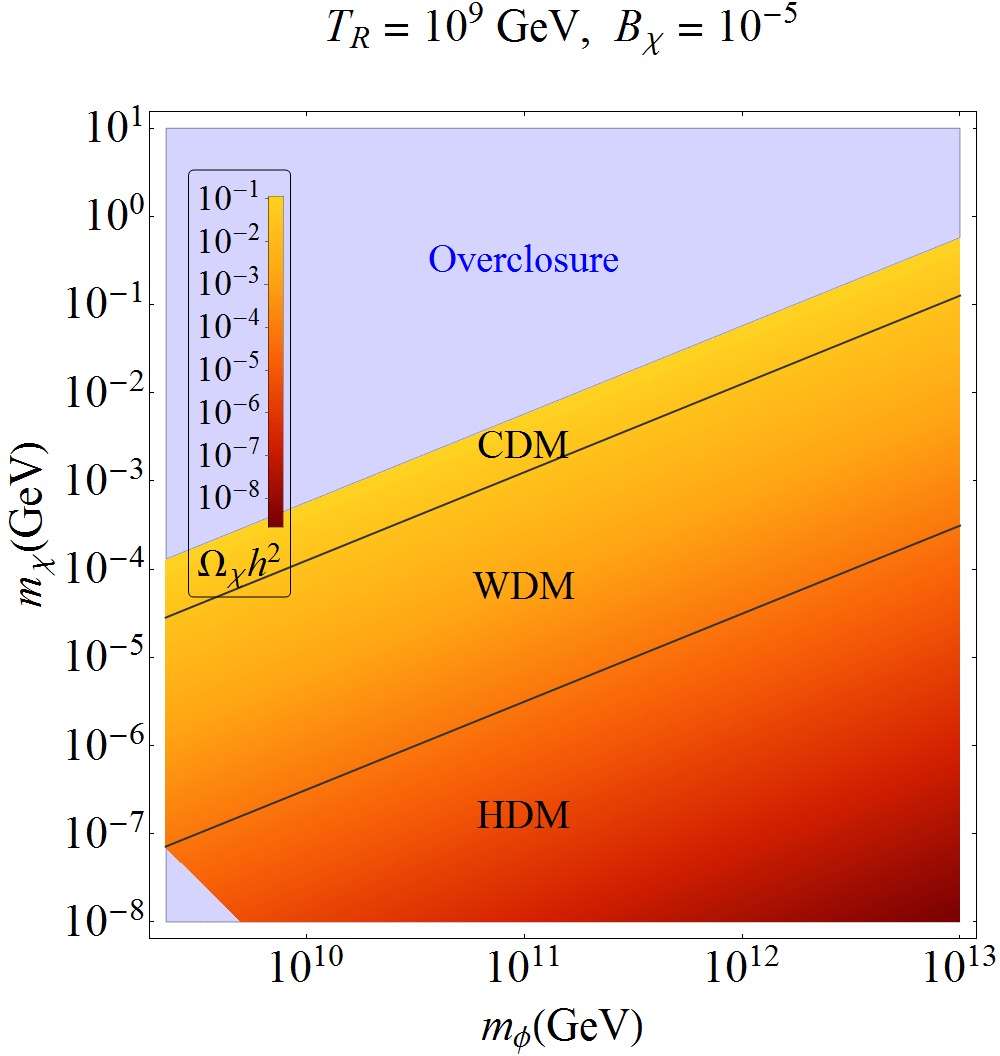}
\includegraphics[width=7.6cm]{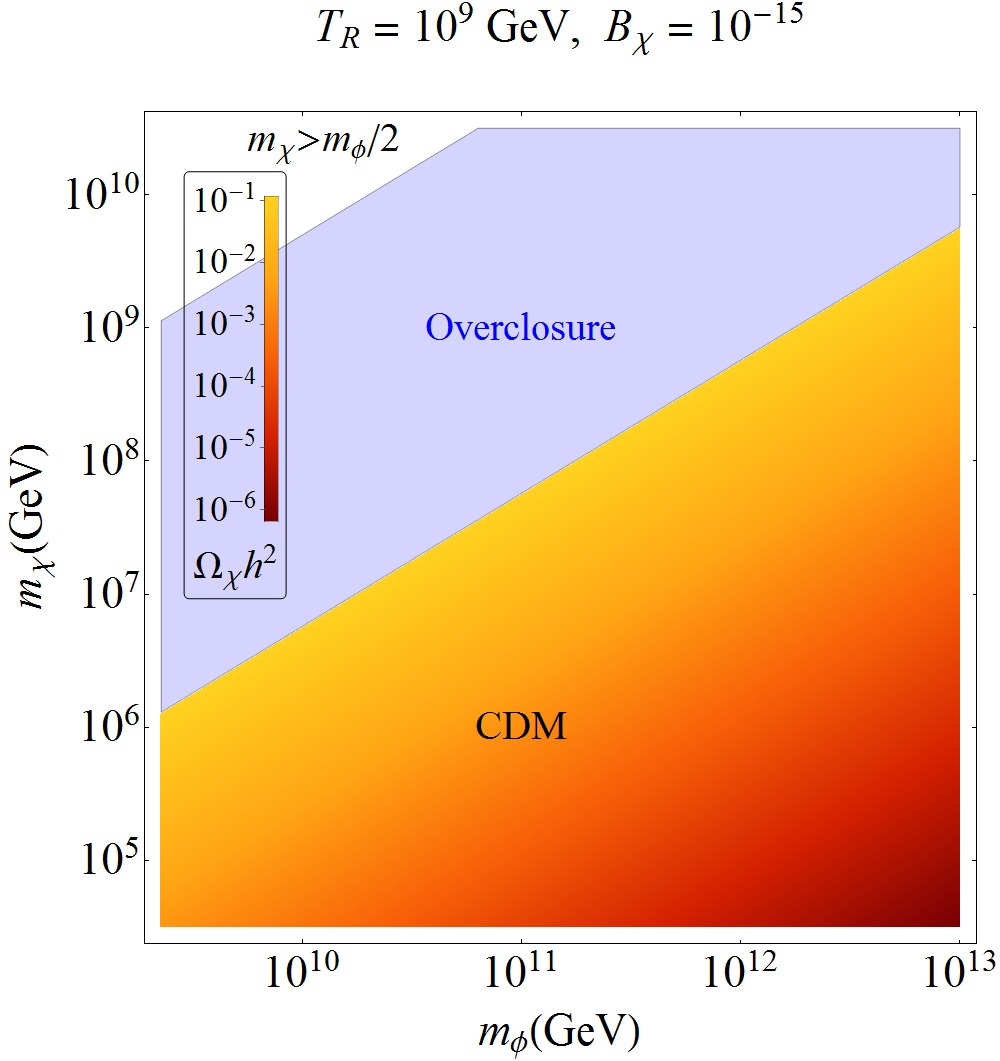}
\caption{The color-coded contours show the relic density of non-thermal DM produced from inflaton decay as a function of the inflaton and DM masses for a fixed reheated temperature $T_R=10^9$ GeV and fixed inflaton branching ratios $B_\chi=10^{-5}$ and $10^{-15}$. We identify the cold, warm and hot DM regions in each case by assuming that the 
corresponding free-streaming length given by Eq.~(\ref{ntlam}) should be $<10$ kpc, between 10 kpc - 2 Mpc, and above 2 Mpc respectively. The blue-shaded region for the CDM case is excluded by the overclosure constraint, as discussed in Section~\ref{sec:5.1}. The additional blue-shaded region in the HDM case is ruled out by the dark radiation limit, as discussed in Section~\ref{sec:5.4}.} \label{fig:6}
\end{figure}

\begin{figure}[t!]
\centering
\includegraphics[width=7.6cm]{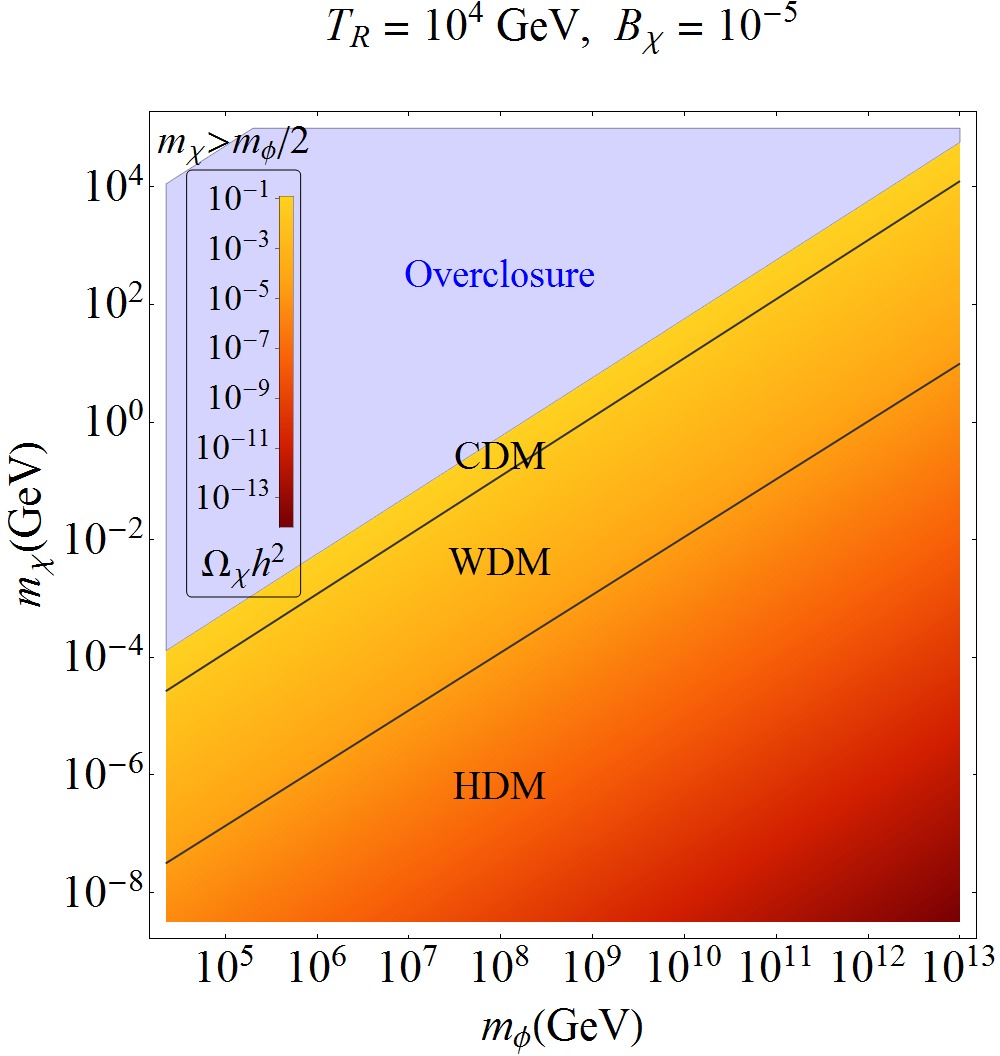}
\includegraphics[width=7.6cm]{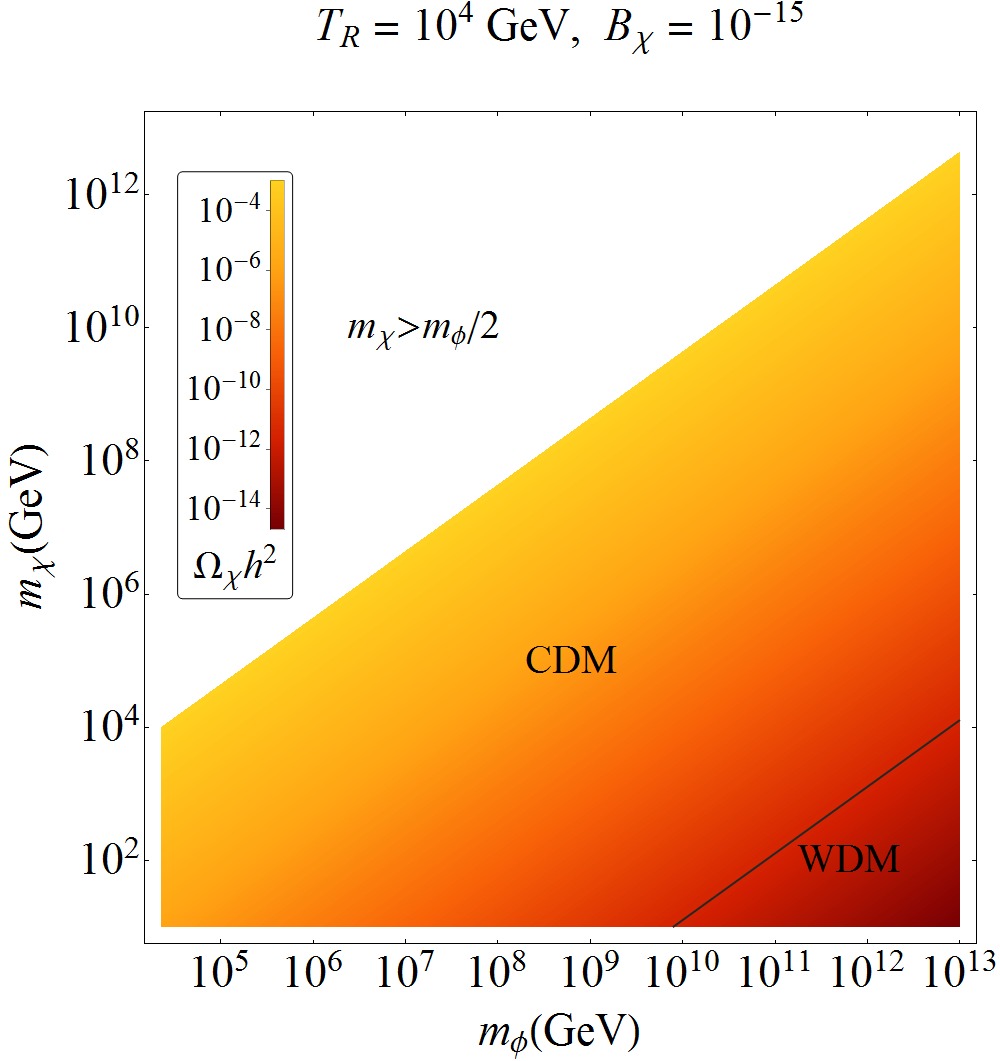}
\caption{The labels are the same as in  Figure~\ref{fig:6}. Here $T_R=10^4$ GeV.}\label{fig:7}
\end{figure}

Now we move on to discuss the non-thermal DM region (green-shaded) in Figures~\ref{fig:scan1}-\ref{fig:scan2b}. As already discussed at length in Section~\ref{sec:non-thermal}, the final relic density of these DM particles is solely determined by the initial conditions, which in our case, are set by the inflaton and DM masses, the reheat temperature and the branching ratio of the inflaton decay to DM. For fixed reheat temperature and inflaton branching ratio, we show in Figures~\ref{fig:6} and \ref{fig:7} the contours for relic density computed using Eq.~(\ref{eq:omega4}) in the $(m_\phi,m_\chi)$ plane. We also calculate the comoving free-streaming length using Eq.~(\ref{ntlam}), and identify the regions with $\lambda_{\rm fs}<10$ kpc as cold DM (CDM), with $\lambda_{\rm fs}>2$ Mpc as hot DM (HDM), and the rest as warm DM (WDM). Note that there is no well-defined boundary between these regions, and we have just chosen some typical values derived from various astrophysical data~\cite{Viel:2013fqw, Boyanovsky:2007ba} for our illustration purposes. We find that the observed DM relic density can be satisfied for a narrow parameter space in the CDM region (the boundary between the blue and orange regions), and the region above this is excluded due to overclosure constraints. For the HDM case with $T_R=10^9$ GeV and $B_\chi=10^{-5}$ (Figure~\ref{fig:6}, left panel), an additional portion of the parameter space (blue-shaded region at bottom-left corner) is ruled out due to the dark radiation constraint, as discussed in Section~\ref{sec:5.4}. 

Similar to the thermal DM case, additional constraints can be derived for specific non-thermal DM candidates. For instance, a popular class of such candidates, known as the Weakly Interacting sub-eV particles (WISPs) such as axions and axion-like particles which often arise as the Nambu-Goldstone bosons associated with some global symmetry breaking, can be constrained from various low-energy experiments involving lasers, microwave and optical cavities, strong electromagnetic fields or torsion balances~\cite{Jaeckel:2010ni}. 
\section{Conclusion} \label{sec:conclusion}
In this paper, we have investigated the thermal and non-thermal properties of DM 
from inflaton decay in a model-independent manner, assuming that the reheating and thermalization of the ambient plasma have happened instantly at a given unique temperature. In the thermal DM scenario, the relic abundance of the DM species is determined by the freeze-out abundance, irrespective of the initial conditions or production mechanism, provided its interaction with the thermal bath is large enough to bring it into LTE soon after its production. For smaller interaction rates when the DM does not attain full LTE, but can still be produced from the thermal bath, one can also obtain the correct relic density through freeze-in mechanism. On the other hand, if the interaction rate is negligibly small so that the DM remains decoupled from the thermal bath from the beginning, the relic density is essentially determined by the initial conditions. Assuming that the DM has a non-zero coupling to the inflaton so that it can be directly produced from the inflaton decay, we have investigated all the above scenarios by tracking the evolution of the DM species from the very onset of its production. We have numerically solved the Boltzmann 
equation for DM number density, and have shown that the inflaton decay to DM inevitably leads to an overclosure of the Universe for a large range of parameter space, especially for non-thermal DM scenarios. This is an important constraint for hidden sector DM models with an arbitrary DM coupling to the inflaton. 
For the thermal DM scenario with large annihilation rates, we show the complementary constraints on the DM parameter space from various experimental searches. On the other hand, the other viable regions for both thermal and non-thermal DM candidates with very small interaction rate remain mostly unexplored.

\section*{Acknowledgments} 
We thank Celine Boehm for useful discussions, and Chiara Arina and Graciela Gelmini for constructive comments on the manuscript. The work of PSBD and AM is supported by the 
Lancaster-Manchester-Sheffield Consortium for Fundamental Physics under STFC grant ST/J000418/1.




\end{document}